\documentclass[12pt]{article}
\usepackage{amsmath}
\usepackage{graphicx}
\usepackage{enumerate}
\usepackage{url}
\usepackage[
backend=biber,
style=apa,
natbib=true,
]{biblatex}
\usepackage[margin=1in]{geometry}
\usepackage{microtype}
\usepackage{caption}
\usepackage{subcaption}
\usepackage[utf8]{inputenc} % Support for special characters
\usepackage{bm} % Bold math symbols
\usepackage{xcolor} % Package for colours
\usepackage{amssymb} % Fancy math symbols
\usepackage{hyperref}

\hypersetup{%
  hypertexnames=false, % less "guessable", but more robust internal link names
  colorlinks = true, % false: boxed links; true: coloured links
  citecolor = blue,
  urlcolor = blue,
  bookmarks = true % Adds a navigation menu in the pdf reader
}

\begin{document}

\def\spacingset#1{\renewcommand{\baselinestretch}%
{#1}\small\normalsize} \spacingset{1}

%%%%%%%%%%%%%%%%%%%%%%%%%%%%%%%%%%%%%%%%%%%%%%%%%%%%%%%%%%%%%%%%%%%%%%%%%%%%%%

\title{\bf An Efficient Workflow for Modelling High-Dimensional Spatial Extremes}

\author{Silius M. Vandeskog\\
  \small{Department of Mathematics, The Norwegian University of Science and Technology (NTNU)}\\
  and \\
  Sara Martino \\
  \small{Department of Mathematics, The Norwegian University of Science and Technology (NTNU)}\\
  and \\
  Rapha{\"e}l Huser \\
  \small{Statistics program, CEMSE Division, King Abdullah University of Science and Technology (KAUST)}}

\maketitle

\bigskip
\begin{abstract}
  A successful model for high-dimensional spatial extremes should, in principle, be able to describe
  both weakening extremal dependence at increasing levels and changes in the type of extremal
  dependence class as a function of the distance between locations. Furthermore, the model should
  allow for computationally tractable inference using inference methods that efficiently extract
  information from data and that are robust to model misspecification.  In this paper, we
  demonstrate how to fulfil all these requirements by developing a comprehensive methodological
  workflow for efficient Bayesian modelling of high-dimensional spatial extremes using the spatial
  conditional extremes model while performing fast inference with \texttt{R-INLA}. We then propose a
  post hoc adjustment method that results in more robust inference by properly accounting for
  possible model misspecification. The developed methodology is applied for modelling extreme hourly
  precipitation from high-resolution radar data in Norway. Inference is computationally efficient,
  and the resulting model fit successfully captures the main trends in the extremal dependence
  structure of the data. Robustifying the model fit by adjusting for possible misspecification
  further improves model performance.
\end{abstract}

\noindent%
{\it Keywords:} Spatial conditional extremes, Robust Bayesian inference, Computational statistics, \texttt{R-INLA}
\vfill

\newpage
\spacingset{1.1}

\section{Introduction}
\label{sec:introduction}

The effects of climate change and the increasing availability of large and high-quality data sets
has lead to a surge of research on the modelling of spatial extremes \citep[e.g.,][]{
  simpson21_condit_model_spatio_tempor_extrem, vandeskog22_model_sub_precip_extrem_blend,
  opitz18_inla_goes_extrem, richards22_model_extrem_spatial_aggreg_precip,
  koch21_trend_extrem_envir_assoc_sever_u, castro-camilo19_splic_gamma_gener_paret_model,
  shooter19_spatial_condit_extrem_ocean_storm_sever,
  koh21_spatiot_wildf_model_point_proces}. Modelling spatial extremes is challenging for two main
reasons: 1) classical models are often not flexible enough to provide realistic descriptions of
extremal dependence, and 2) inference can be computationally demanding or intractable, so modellers
must often rely on less efficient inference methods; see
\citet{huser22_advan_statis_model_spatial_extrem} for a review of these challenges. In this paper,
we propose a comprehensive methodological workflow, as well as practical strategies, on how to
perform efficient and flexible high-dimensional modelling of spatial extremes.

An important component of spatial extreme value theory is the characterisation of a spatial process'
asymptotic dependence properties \citep[e.g.,][]{coles99_depen_measur_extrem_value_analy}. Two
random variables with a positive limiting probability to experience their extremes simultaneously
are denoted asymptotically dependent. Otherwise, they are denoted asymptotically independent. As
demonstrated by \citet{sibuya60_bivar_extrem_statis}, two asymptotically independent random
variables may still be highly correlated and thus exhibit large amounts of so-called sub-asymptotic
dependence.  Thus, correct estimation of both asymptotic and sub-asymptotic dependence properties is
of utmost importance when assessing the risks of spatial extremes.

Most classical models for spatial extremes are based on max-stable processes
\citep{davison12_statis_model_spatial_extrem, davison19_spatial_extrem}. These allow for rich
modelling of asymptotic dependence, but are often too rigid in their descriptions of asymptotic
independence and sub-asymptotic dependence.  Other approaches have been proposed, such as
scale-mixture models \citep{engelke19_extrem_depen_random_scale_const,
  huser19_model_spatial_proces_with_unknow}, which allow for rich modelling of both asymptotic
dependence and independence, and a more flexible description of sub-asymptotic dependence. However,
these models require that all location pairs share the same asymptotic dependence class, which is
problematic as one would expect neighbouring locations to be asymptotically dependent and far-away
locations to be asymptotically independent.  Max-mixture model
\citep{wadsworth12_depen_model_spatial_extrem} allow for even more flexible modelling of
sub-asymptotic dependence, and for changing the asymptotic dependence class as a function of
distance. However, it is often difficult to estimate the key model parameter, which describes the
transition between extremal dependence classes. Additionally, these models must often rely on less
efficient inference methods. Further improvements are given by the kernel convolution model of
\citet{krupskii22_model_spatial_tail_depen_cauch_convol_proces}, more recent scale-mixture models
such as that of \citet{hazra21_realis_fast_model_spatial_extrem}, and the spatial conditional
extremes model of \citet{wadsworth22_higher_spatial_extrem_singl_condit}, which all allows for
flexible modelling of different extremal dependence classes as a function of distance.  The spatial
conditional extremes model allows for a particularly simple way of modelling spatial extremes.  It
is based on the conditional extremes model of
\citet{heffernan04_condit_approac_multiv_extrem_values,
  heffernan07_limit_laws_random_vector_with_extrem_compon}, which describes the behaviour of a
random vector conditional on one of its components being extreme, and it can be interpreted as a
semi-parametric regression model, which makes it intuitive and simple to tailor or extend. Due to
its high flexibility and conceptual simplicity, this is our chosen model for high-dimensional
spatial extremes.

To make the spatial conditional extremes model computationally efficient in higher dimensions,
\citet{wadsworth22_higher_spatial_extrem_singl_condit} propose to model spatial dependence using a
residual random process constructed from a Gaussian copula and delta-Laplace marginal distributions.
However, inference for Gaussian processes typically requires computing the inverse of the covariance
matrix, whose cost scales cubicly with the model dimension. Thus,
\citet{simpson20_high_model_spatial_spatio_tempor}, propose to exchange the delta-Laplace process
with a Gaussian Markov random field \citep{rue05_gauss_markov_random_field} created using the
so-called stochastic partial differential equations (SPDE) approach of
\citet{lindgren11_explic_link_between_gauss_field}.  Furthermore, in order to perform spatial
high-dimensional Bayesian inference, \citet{simpson20_high_model_spatial_spatio_tempor} modify the
spatial conditional extremes model into a latent Gaussian model, which allows for performing
inference using the integrated nested Laplace approximation
\citep[INLA;][]{rue09_approx_bayes_infer_laten_gauss}, implemented in the \texttt{R-INLA} software
\citep{rue17_bayes_comput_with_inla}. This allows for a considerable improvement in the Bayesian
modelling of high-dimensional spatial extremes. However, there is still much room for
improvement. In this paper, we thus build upon the modelling framework of
\citet{simpson20_high_model_spatial_spatio_tempor} and develop a more general methodology for
modelling spatial conditional extremes with \texttt{R-INLA}. We also point out a theoretical
weakness in the constraining methods proposed by
\citet{wadsworth22_higher_spatial_extrem_singl_condit} and used by
\citet{simpson20_high_model_spatial_spatio_tempor}, and we demonstrate a computationally efficient
way of fixing it.

As most statistical models for extremes are based on asymptotic arguments and assumptions, a certain
degree of misspecification will always be present when modelling finite amounts of
data. Additionally, model choices made for reasons of computational efficiency, such as adding
Markov assumptions to a spatial random field, may lead to further misspecification.
%the extent of which is expected to increase with larger and more complex data sets.
This complicates Bayesian inference and can result in misleading posterior distributions
\citep{ribatet12_bayes_infer_from_compos_likel, kleijn12_berns_von_mises_theor_missp}.  One should
therefore strive to make inference more robust towards misspecification when modelling
high-dimensional spatial extremes. \citet{shaby14_open_faced_sandw_adjus_mcmc} proposes a method for
more robust inference through a post hoc transformation of posterior samples created using Markov
chain Monte Carlo (MCMC) methods. Here, we develop a refined version of his adjustment method, and
we use it for performing more robust inference with \texttt{R-INLA}.

As extreme behaviour is, by definition, rare, inference with the conditional extremes model often
relies on a composite likelihood that combines data from different conditioning sites under the
working assumption of independence \citep{heffernan04_condit_approac_multiv_extrem_values,
  wadsworth22_higher_spatial_extrem_singl_condit, richards22_model_extrem_spatial_aggreg_precip,
  simpson21_condit_model_spatio_tempor_extrem}. However, composite likelihoods can lead to large
amounts of misspecification \citep{ribatet12_bayes_infer_from_compos_likel}, and
\citet{simpson20_high_model_spatial_spatio_tempor} thus abstain from using a composite likelihood to
avoid the problems that occur when performing Bayesian inference with a composite likelihood using
\texttt{R-INLA}. We show that the post hoc adjustment method accounts for the misspecification from
the composite likelihood, thus allowing for more efficient inference using considerably more data.

To sum up, in this paper we develop a general workflow for performing high-dimensional modelling of
spatial extremes using the spatial conditional extremes model. We improve upon the work of
%\citet{wadsworth22_higher_spatial_extrem_singl_condit} and
\citet{simpson20_high_model_spatial_spatio_tempor} by developing a more general, flexible and
computationally efficient methodology for modelling spatial conditional extremes with
\texttt{R-INLA} and the SPDE approach. Then, we make inference more robust towards misspecification
by extending the post hoc adjustment method of \citet{shaby14_open_faced_sandw_adjus_mcmc}, and we
further apply this adjustment method for more efficient inference by combining information from
multiple conditioning sites.

The remainder of the paper is organised as follows: In Section~\ref{sec:conditional}, the spatial
conditional extremes model is presented as a flexible choice for modelling spatial
extremes. Modifications and assumptions that allow for computationally efficient inference with
improved data utilisation are also presented.  Then, in Section~\ref{sec:implementation}, we develop a
general methodology for implementing a large variety of spatial conditional extremes models in
\texttt{R-INLA}. Section~\ref{sec:robustifying-inference} examines the problems that can occur when
performing Bayesian inference based on a misspecified likelihood, and demonstrates how to perform
more robust inference with \texttt{R-INLA} by accounting for possible misspecification. In
Section~\ref{sec:simulation}, a simulation study is presented where we demonstrate and validate our
entire workflow for high-dimensional modelling of spatial extremes. Then, in
Section~\ref{sec:case-study}, our proposed workflow is applied for modelling extreme hourly
precipitation from high-resolution radar data in Norway. Finally, we conclude in
Section~\ref{sec:conclusion} with some discussion and perspectives on future research.

\section{Flexible modelling with spatial conditional extremes}
\label{sec:conditional}

\subsection{The spatial conditional extremes model}
\label{sec:conditional-prerequisites}

Let \(Y(\bm s)\) be a random process defined over space
\((\bm s \in \mathcal S \subset \mathbb R^2)\) with Laplace margins. For this random process,
\citet{wadsworth22_higher_spatial_extrem_singl_condit} assume the existence of standardising
functions \(a(\bm s; \bm s_0, y_0)\) and \(b(\bm s; \bm s_0, y_0)\) such that, for a large enough
threshold \(t\),
\begin{equation}
  \label{eq:conditional-model}
  \left[Y(\bm s) \mid Y(\bm s_0) = y_0 > t \right] \overset{d}{=} a(\bm s; \bm s_0, y_0) + b(\bm
  s; \bm s_0, y_0) Z(\bm s; \bm s_0), \quad \bm s, \bm s_0 \in \mathcal S,
\end{equation}
where \(Z(\bm s; \bm s_0)\) is a random process satisfying \(Z(\bm s_0; \bm s_0) = 0\) almost
surely, and \(a(\bm s; \bm s_0, y_0) \leq y_0\), with equality when \(\bm s = \bm s_0\).
The degree of asymptotic dependence may be measured through the extremal correlation coefficient
\[
  \chi(\bm s_1, \bm s_2) = \lim_{p \rightarrow 1} \chi_p(\bm s_1, \bm s_2) = \lim_{p \rightarrow 1}
  \text{P}(Y(\bm s_1) > F^{-1}_Y(p) \mid Y(\bm s_2) > F^{-1}_Y(p)),
\]
where \(F^{-1}_Y(p)\) is the marginal quantile function of the process \(Y(\bm s)\).
If \(\chi(\bm s_1, \bm s_2) > 0\), then \(Y(\bm s_1)\) and \(Y(\bm s_2)\) are asymptotically
dependent, whereas if \(\chi(\bm s_1, \bm s_2) = 0\), they are asymptotically independent.
It is known that
\(Y(\bm s)\) and \(Y(\bm s_0)\), defined in~\eqref{eq:conditional-model}, are asymptotically
dependent when \(a(\bm s; \bm s_0, y_0) = y_0\) and \(b(\bm s; \bm s_0, y_0) = 1\), while they are
asymptotically independent when \(a(\bm s; \bm s_0, y_0) < y_0\)
\citep{heffernan04_condit_approac_multiv_extrem_values}. However, under asymptotic independence, the
convergence of \(\chi_p(\cdot)\) to \(\chi(\cdot)\) is slower for larger values of \(a(\cdot)\) and
\(b(\cdot)\).
% which therefore corresponds to larger degrees of sub-asymptotic dependence.

\citet{wadsworth22_higher_spatial_extrem_singl_condit} provide some guidance on parametric functions
for \(a(\cdot)\) and \(b(\cdot)\) together with parametric distributions for \(Z(\cdot)\) that cover
a large range of already existing models. For modelling \(a(\cdot)\) they specifically propose the
parametric function
%\begin{equation}
%  \label{eq:a_wadsworth}
%  a(\bm s; \bm s_0, y_0) = \alpha(\|\bm s - \bm s_0\|) y_0 =
%  \begin{cases}
%    y_0
%    & \|\bm s - \bm s_0\| < \Delta \\
%    y_0 \exp\left\{-\left[(\|\bm s - \bm s_0\| - \Delta) / \lambda_a\right]^{\kappa_a}\right\}
%    & \|\bm s - \bm s_0\| \geq \Delta
%  \end{cases},
%\end{equation}
\begin{equation}
  \label{eq:a_wadsworth}
  a(\bm s; \bm s_0, y_0) = y_0 \alpha(\|\bm s - \bm s_0\|) = y_0 \exp\left\{-\left[\max(0, \|\bm s -
      \bm s_0\| - \Delta) / \lambda_a\right]^{\kappa_a}\right\},
  % \quad d = \|\bm s - \bm s_0\|,
\end{equation}
with parameters \(\Delta \geq 0\) and \(\lambda_a, \kappa_a > 0\). This function yields a model with
asymptotic dependence for locations closer to the conditioning site than a distance \(\Delta\),
while displaying asymptotic independence for distances larger than \(\Delta\), with a weakening
sub-asymptotic dependence as we move further away from \(\bm s_0\).  To the best of our knowledge,
this model (and its sub-models) has been adopted by a majority of spatial conditional extremes
modellers.  Several forms are proposed for \(b(\cdot)\), including the form
\(b(\bm s; \bm s_0, y_0) = y_0^\beta\), when \(\Delta = 0\).  This allows for modelling asymptotic
independence with positive dependence, with the \(\beta\) parameter helping to control the speed of
convergence of \(\chi_p(\bm s_1, \bm s_2)\) to \(\chi(\bm s_1, \bm s_2)\).  A weakness of this form
is that it enforces the same positive dependence for all distances, including large distances where
the observations should be independent of \(Y(\bm s_0)\). To remedy this issue,
\citet{wadsworth22_higher_spatial_extrem_singl_condit} also propose the model
\(b(\bm s; \bm s_0, y_0) = 1 + a(\bm s; \bm s_0, y_0)^\beta\), which converges to one as the
distance increases. Alternatively, \citet{shooter21_basin_spatial_condit_extrem_sever_ocean_storm}
and \citet{richards22_model_extrem_spatial_aggreg_precip} have proposed different models on the form
\(b(\bm s; \bm s_0, y_0) = y_0^{\beta(\|\bm s - \bm s_0\|)}\), where they let the function
\(\beta(d)\) converge to zero as the distance \(d \rightarrow \infty\).

%the best model for the standardising functions \(a(\cdot)\) and \(b(\cdot)\) depends on the application.
%to application.

Clearly, the best model for the standardising functions \(a(\cdot)\) and \(b(\cdot)\) depends
on the application. Therefore, in Section~\ref{sec:implementation}, we develop a general
methodology for implementing the conditional spatial extremes model in \texttt{R-INLA} for any kind
of functions \(a(\cdot)\) and \(b(\cdot)\). In addition, we provide practical guidance and
diagnostics for selecting appropriate standardising functions in our simulation study in
Section~\ref{sec:simulation} and data application in Section~\ref{sec:case-study}.

\subsection{Modifications for high-dimensional modelling}
\label{sec:model_for_high-dim}

To perform high-dimensional inference,
\citet{wadsworth22_higher_spatial_extrem_singl_condit} propose to model \(Z(\cdot)\) as a random
process with a Gaussian copula and delta-Laplace marginal distributions. Their proposed model for
\(Z(\cdot)\) has later seen usage by, e.g., \citet{shooter21_spatial_depen_extrem_seas_north,
  shooter21_basin_spatial_condit_extrem_sever_ocean_storm,
  shooter22_multiv_spatial_condit_extrem_extrem_ocean_envir} and
\citet{richards22_model_extrem_spatial_aggreg_precip}. However, in order to perform Bayesian
inference with \texttt{R-INLA},
%stricter modelling assumptions are necessary. Thus,
\citet{simpson20_high_model_spatial_spatio_tempor} modify \eqref{eq:conditional-model} into a latent
Gaussian model by adding a Gaussian nugget effect and requiring \(Z(\cdot)\) to be a fully Gaussian
random field. This gives the model
\begin{equation}
  \label{eq:conditional-model-simpson}
  \left[Y(\bm s) \mid Y(\bm s_0) = y_0 > t \right] \overset{d}{=} a(\bm s; \bm s_0, y_0) + b(\bm s; \bm
  s_0, y_0) Z(\bm s; \bm s_0) + \epsilon(\bm s; \bm s_0), 
\end{equation}
where \(\epsilon(\bm s; \bm s_0)\) is Gaussian white noise with constant variance, satisfying
\(\epsilon(\bm s_0; \bm s_0) = 0\) almost surely. They further assume that \(Z(\cdot)\) has zero
mean and a Matérn covariance structure, so that it can be approximated using the SPDE approach of
\citet{lindgren11_explic_link_between_gauss_field}, which speeds up inference by approximating the
precision matrix of \(Z(\cdot)\) with a sparse and low-rank matrix.  However, making the precision
matrix too sparse and/or low-rank leads to some model misspecification, which is further amplified
by the fact that \(Y(\bm s)\) has Laplace marginal distributions, but is modelled using a fully
Gaussian random field. We here nevertheless adopt the modelling assumptions of
\citet{simpson20_high_model_spatial_spatio_tempor}, as we find them necessary for performing truly
high-dimensional Bayesian inference with \texttt{R-INLA}. However, unlike
\citet{simpson20_high_model_spatial_spatio_tempor}, we then account for the possible
misspecification of these assumptions using the robustifying approach described in
Section~\ref{sec:robustifying-inference}.

\subsection{Efficient data utilisation with a composite likelihood}
\label{sec:composite_likelihood}

The spatial conditional extremes model consists in modelling a spatial process conditional on
extreme behaviour at a predefined conditioning site.  However, inference is often made challenging
because the conditioning site contains few observed extremes. To strengthen inference it is
therefore common to assume stationarity, in the sense that all parameters of \(a(\cdot)\),
\(b(\cdot)\), \(Z(\cdot)\) and \(\epsilon(\cdot)\) are independent of the conditioning site. Under
such stationarity, it is possible to combine information from multiple conditioning sites into one
global model fit, using the composite likelihood of
\citet{heffernan04_condit_approac_multiv_extrem_values} and
\citet{wadsworth22_higher_spatial_extrem_singl_condit}. Given observations
\(\mathcal Y = \{y_i(\bm s_j): i = 1, 2, \ldots, n, j = 1, 2, \ldots, m\}\) from \(n\) time points
and \(m\) locations, the composite log-likelihood may be expressed as
\begin{equation}
  \label{eq:conditional-composite-likelihood}
  \ell_c(\bm \theta; \mathcal Y) = 
  \sum_{i = 1}^n \sum_{j = 1}^m \ell(\bm \theta; \bm y_{i, -j} \mid y_i(\bm s_j)) I(y_i(\bm s_j) > t),
\end{equation}
where \(\ell(\cdot)\) is the log-likelihood of the conditional extremes model, \(\bm y_{i, -j}\) is
a \((m - 1)\)-dimensional vector containing observations from time point \(i\), for all locations
except \(\bm s_j\), \(I(\cdot)\) is the indicator function and \(\bm \theta\) contains all
parameters of the spatial conditional extremes model. If \(m\) is too large, one may choose to build
the composite likelihood using only a subset of the available conditioning sites, and
\(\bm y_{i, -j}\) may be modified to contain only a subset of the available observations from time
point \(i\), which may vary with both \(i\) and \(j\).

The composite likelihood is not a valid likelihood, since multiple of the terms
in~\eqref{eq:conditional-composite-likelihood} may contain the same observations. Incorrectly
interpreting the composite likelihood as a true likelihood is therefore tantamount to specifying a
model in which \([\bm y_{i, -j} \mid y_i(\bm s_j)]\) is (wrongly) assumed to be independent from
\([\bm y_{i, -k} \mid y_i(\bm s_k)]\) for all time points \(i\) and locations pairs
\((\bm s_j,\bm s_k)\) with \(y_i(\bm s_j) > t\) and \(y_i(\bm s_k) > t\). Performing inference with
the composite likelihood can therefore lead to considerable misspecification, which should be
accounted for before drawing conclusions from the model fit. In
Section~\ref{sec:robustifying-inference}, we therefore show how to robustify inference by accounting
for the possible model misspecification. To the best of our knowledge, our paper is the first
attempt to perform Bayesian inference for the spatial conditional extremes model based on a
composite likelihood.

\section{Fast inference using \texttt{R-INLA}}
\label{sec:implementation}

\subsection{Latent Gaussian model framework}

\texttt{R-INLA} performs inference on latent Gaussian models of the form
\[
  \begin{aligned}
    \left[y_i \mid \bm u, \bm \theta_1 \right]& \overset{i.i.d.}{\sim} \mathcal \pi(y_i \mid
     \eta_i(\bm u), \bm \theta_1),\ i = 1, 2, \ldots, n, \\
    \left[\bm u \mid \bm \theta_2\right]& \sim \mathcal N(\bm \mu(\bm \theta_2), \bm Q^{-1}(\bm \theta_2)), \\
    \left(\bm \theta_1^{\top}, \bm \theta_2^{\top}\right)^{\top}& \sim \pi(\bm \theta_1) \pi(\bm \theta_2),
  \end{aligned}
\]
where \(\bm u\) is a latent Gaussian field with mean \(\bm \mu(\bm \theta_2)\) and precision matrix
\(\bm Q(\bm \theta_2)\), and the hyperparameters
\(\bm \theta = (\bm \theta_1^{\top}, \bm \theta_2^{\top})^{\top}\) are assigned priors
\(\pi(\bm \theta_1)\) and \(\pi(\bm \theta_2)\).  Observations $\bm y = (y_1, \ldots, y_n)^{\top}$
are linked to the latent field through the linear predictor
\(\bm \eta = (\eta_1(\bm u), \ldots, \eta_n(\bm u))^{\top} = \bm A \bm u\),
%\(\bm \eta(\bm u) = \bm A \bm u \),
where \(\bm A\) is a known design matrix. This linear predictor defines the location parameter of
the likelihood \(\pi(\bm y \mid \bm \eta, \bm \theta_1)\), via a possibly non-linear link
function. All observations are assumed to be conditionally independent given \(\bm \eta\) and
\(\bm \theta_1\), so that
\(\pi(\bm y\mid \bm \eta, \bm \theta_1) = \prod_{i = 1}^n\pi( y_i\mid \eta_i(\bm u), \bm
\theta_1)\).  For computational reasons, when using \texttt{R-INLA}, the likelihood must be chosen
from a predefined set of likelihood functions. The linear predictor can be decomposed into
\(N \geq 1\) components, \(\bm \eta = \bm A^{(1)} \bm u^{(1)} + \dots + \bm A^{(N)}\bm u^{(N)}\),
where each component represents, e.g., an intercept term, a linear combination of regression
coefficients, an SPDE component, etc. All of these components must either be predefined in
\texttt{R-INLA} or defined by the user, using the \texttt{rgeneric} framework or the recently added
\texttt{cgeneric} framework.

The spatial conditional extremes model in~\eqref{eq:conditional-model-simpson} corresponds to a
latent Gaussian model where the likelihood is Gaussian with variance \(\theta_1\), say, and the
linear predictor is equal to \(a(\bm s; \bm s_0, y_0) + b(\bm s; \bm s_0, y_0) Z(\bm s; \bm s_0)\),
with \(\bm \theta_2\) containing the parameters of \(a(\cdot)\), \(b(\cdot)\) and \(Z(\cdot)\).
%Here, we define the linear predictor as a sum of the two latent components
%\(a(\bm s; \bm s_0, y_0)\) and \(b(\bm s; \bm s_0, y_0) Z(\bm s; \bm s_0)\).
For most forms of \(a(\cdot)\) and \(b(\cdot)\), \texttt{R-INLA} does not contain suitable
predefined model components for describing the linear predictor, so we must define these manually.
In order to define a new \texttt{R-INLA} component \(\bm u^{(N + 1)}\), with parameters
\(\bm \theta^{(N + 1)}\), using one of the \texttt{rgeneric}/\texttt{cgeneric} frameworks, one must
provide functions written in \texttt{R} of \texttt{C}, respectively, that compute the precision
matrix mean and prior density of \(\bm u^{(N + 1)}\) for any value of \(\bm \theta^{(N + 1)}\). The
\texttt{cgeneric} framework yields considerably faster inference than the \texttt{rgeneric}
framework, but it requires knowledge of the lower-level \texttt{C} programming language. In this
paper, we propose a method for defining model components for \(a(\bm s; \bm s_0, y_0)\) and
\(b(\bm s; \bm s_0, y_0) Z(\bm s; \bm s_0)\) using the \texttt{rgeneric}/\texttt{cgeneric}
frameworks, for any kind of functions \(a(\cdot)\) and \(b(\cdot)\). In the online supplementary
material, we provide the necessary code for defining the models used in Section~\ref{sec:simulation}
and~\ref{sec:case-study} with the \texttt{cgeneric} framework.

\subsection{Defining \texorpdfstring{\(b(\bm s; \bm s_0, y_0) Z(\bm s; \bm s_0)\)}{b Z} in \texttt{R-INLA}}
\label{sec:implementing_bZ}

%We propose a method for defining the constrained, non-stationary field
%\(b(\bm s; \bm s_0, y_0) Z(\bm s; \bm s_0)\) with the \texttt{rgeneric}/\texttt{cgeneric}
%frameworks. Here, we first show how to define an unconstrained non-stationary random
%field. Then, we propose a method for adding constraints to an already defined random field.

The SPDE approach creates a Gaussian Markov random field \(\widehat Z(\bm s)\) that is an
approximation to a Gaussian random field \(Z(\bm s)\) with Matérn covariance function
\begin{equation}
  \label{eq:matern}
  \text{Cov}(Z(\bm s), Z(\bm s')) = \frac{\sigma^2}{2^{\nu - 1}\Gamma(\nu)}(\kappa \|\bm s - \bm
  s'\|)^\nu K_\nu(\kappa \|\bm s - \bm s'\|),
\end{equation}
where \(\sigma^2\) is the marginal variance, \(\nu > 0\) is a smoothness parameter,
\(\rho = \sqrt{8 \nu} / \kappa\) is a range parameter and \(K_\nu\) is the modified Bessel function
of the second kind and order \(\nu\). The smoothness parameter \(\nu\) is difficult to estimate from
data and is therefore often given a fixed value \citep{lindgren15_bayes_spatial_model_r_inla}. The
SPDE approximation \(\widehat Z(\bm s)\) is constructed as a linear combination of Gaussian Markov
random variables on a triangulated mesh, i.e.,
\(\widehat Z(\bm s) = \sum_{i = 1}^M \phi_i(\bm s) W_i\),
%\[
%  \widehat Z(\bm s) = \sum_{i = 1}^M \phi_i(\bm s) W_i \overset{d}{\approx} Z(\bm s),
%\]
where \(W_1, \ldots, W_M\) are random variables from a Gaussian Markov random field, and
\(\phi_i, \ldots, \phi_M\) are piecewise linear basis functions. In order to approximate the
non-stationary Gaussian random field \(b(\bm s; \bm s_0, y_0) Z(\bm s)\) with the SPDE approach, for
any function \(b(\bm s; \bm s_0, y_0)\), we modify the weights \(W_i\) to get
\begin{equation}
  \label{eq:spde-approx}
  \widehat Z_b(\bm s; \bm s_0, y_0) = \sum_{i = 1}^M \phi_i(\bm s) b(\bm s_i; \bm s_0, y_0) W_i
\end{equation}
where \(\bm s_1, \ldots \bm s_M\) are the locations of the \(M\) mesh nodes. This shares some
similarities with the approach of
\citet{ingebrigtsen14_spatial_model_with_explan_variab_depen_struc} for implementing non-stationary
SPDE fields. Since \(\widehat Z_b(\cdot)\) is a linear combination of Gaussian random variables,
its variance equals
\[
  \text{Var}\left(\widehat Z_b(\bm s; \bm s_0, y_0)\right) = \sum_{i, j = 1}^M \phi_i(\bm s) \phi_j(\bm
  s) b(\bm s_i; \bm s_0, y_0) b(\bm s_j; \bm s_0, y_0) \text{Cov}\left(W_i, W_j\right),
\]
which is unequal to \(b(\bm s; \bm s_0, y_0)^2 \sigma^2\), the variance of
\(b(\bm s; \bm s_0, y_0) Z(\bm s)\). However, if \(\bm s\) coincides with a mesh node, then one of
the basis functions equals \(1\), while the others equal \(0\), giving
\(\text{Var}\left(\widehat Z_b(\bm s; \bm s_0, y_0)\right) = b(\bm s; \bm s_0, y_0)^2
\text{Var}(W_i)\), which is much closer to the correct variance. On the contrary, if \(\bm s\) is
far away from a mesh node, the variance of \(\widehat Z_b(\bm s; \bm s_0, y_0)\) may be considerably
different from \(b(\bm s; \bm s_0, y_0)^2 \sigma^2\). If possible, it is therefore recommended to
use a fine mesh, so all observation locations are close enough to a mesh node.

%Thus, if the mesh used for defining \(\widehat Z_b(\cdot)\) is
%dense and all data locations in \(\mathcal S\) are close to a mesh node, \(\widehat Z_b(\cdot)\) can
%provide a good approximation to \(b(\bm s; \bm s_0, y_0) Z(\bm s)\), but if the mesh is too coarse,
%the approximation may deteriorate. It is therefore recommended to consider a fine
%mesh, if possible.

The process \(\widehat Z_b(\cdot)\) approximates the unconstrained Gaussian random field
\(b(\bm s; \bm s_0, y_0) Z(\bm s)\). However, in order to define the conditional extremes model in
\texttt{R-INLA}, we need to approximate the constrained field
\(b(\bm s; \bm s_0, y_0) Z(\bm s; \bm s_0)\), where \(Z(\bm s_0; \bm s_0) = 0\) almost surely.
\citet{wadsworth22_higher_spatial_extrem_singl_condit} describe two different methods for turning an
unconstrained Gaussian field \(Z(\bm s)\) into a constrained field \(Z(\bm s; \bm s_0)\). The first
one is to constrain the field by conditioning, i.e.,
\(Z(\bm s; \bm s_0) = [Z(\bm s) \mid Z(\bm s) = 0]\); and the second one is to constrain it by
subtraction, i.e., \(Z(\bm s; \bm s_0) = Z(\bm s) - Z(\bm s_0)\).  In their case studies,
\citet{wadsworth22_higher_spatial_extrem_singl_condit} use the first method, while
\citet{simpson20_high_model_spatial_spatio_tempor} use the second method. We argue that constraining
by subtraction yields unrealistic dependence structures, and should be avoided if other alternatives
are available. A quick computation indeed shows that if \(Z(\bm s; \bm s_0)\) is a stationary random
process that has been constrained through subtraction, 
then the limiting correlation between \(Z(c \bm s; \bm s_0)\) and \(Z(-c \bm s; \bm s_0)\), as \(c
\rightarrow \infty\) equals \(1/2\). Furthermore, the limiting correlation of \(Z(\bm s_0 + \Delta
\bm s; \bm s_0)\) and \(Z(\bm s_0 - \Delta \bm s; \bm s_0)\) as \(\|\Delta \bm s\| \rightarrow 0\)
is often negative and equals 0 if the unconstrained random field had an exponential correlation
function or \(-1\) if the unconstrained field had a Gaussian correlation function.
%\begin{equation}
%  \label{eq:constraining_Z}
%    \lim_{c \rightarrow \infty} \text{Corr}(Z(c \bm s; \bm s_0), Z(-c \bm s; \bm s_0)) =
%    \frac{1}{2},
%\end{equation}
%and if \(Z(\bm s)\) is isotropic with autocorrelation 
%\(\text{Corr}(Z(\bm s), Z(\bm s')) = r(\|\bm s - \bm s'\|)\), then
%\begin{equation}
%  \label{eq:constraining_Z_2}
%    \text{Corr}(Z(\bm s_0 + \Delta \bm s; \bm s_0), Z(\bm s_0
%      - \Delta \bm s; \bm s_0)) =
%      \frac{1 - 2 r(\|\Delta \bm s\|) + r(2\|\Delta \bm s\|)}{2(1 - r(\|\Delta \bm s\|))}.
%\end{equation}
%The limit of \eqref{eq:constraining_Z_2} as \(\|\Delta \bm s\| \rightarrow 0\) is often negative,
%and equals \(0\) for the exponential autocorrelation function and \(-1\) for the Gaussian
%autocorrelation function.
Thus, with the subtraction approach, points that are infinitely far away
from each other are strongly correlated while points that are infinitesimally close to each other
might be negatively correlated or independent.

\texttt{R-INLA} contains an implementation for constraining a random field by conditioning on linear
combinations of itself \citep{rue09_approx_bayes_infer_laten_gauss}. One can therefore easily
constrain \(\widehat Z_b(\cdot)\) by conditioning using the \texttt{extraconstr} option in
\texttt{R-INLA}. Unfortunately, this conditioning method requires the computation of an
\((n \times k)\)-dimensional dense matrix, where \(n\) is the number of rows of the precision matrix
and \(k\) is the number of added constraints.  In practice, we therefore experience that
constraining by conditioning with \texttt{R-INLA} requires considerably more computational
resources, and that it quickly turns intractable for large data sets.

We propose a third method for constraining the residual field with \texttt{R-INLA}. It is known
that, for a Gaussian random vector \(\bm y = (\bm y_1^{\top}, \bm y_2^{\top})^{\top}\) with zero
mean and precision matrix
\[
\bm Q = \begin{pmatrix} \bm Q_{11} & \bm Q_{12} \\ \bm Q_{21} & \bm Q_{22} \end{pmatrix},
\]
the
conditional distribution of \([\bm y_1 \mid \bm y_2 = \bm 0]\) is Gaussian with zero mean and
precision matrix \(\bm Q_{11}\) \citep{rue05_gauss_markov_random_field}. Thus, if we ensure that a
mesh node coincides with \(\bm s_0\), we can constrain \(Z(\cdot)\) by removing all rows and columns
of \(\bm Q\) that correspond to the mesh node at \(\bm s_0\). This is easily achievable using the
\texttt{rgeneric}/\texttt{cgeneric} framework, and it requires no extra computational effort.

% A third method for constraining the residual field is used by
% \citet{richards22_model_extrem_spatial_aggreg_precip}. They model the variance of the residual field
% as a function of the distance \(\|\bm s - \bm s_0\|\) such that the variance is zero when
% \(\|\bm s - \bm s_0\| = 0\). This can easily be implemented by modifying the standardising function
% \(b(\cdot)\) in \(\widehat Z_b(\cdot)\) such that \(b(\bm s_0; \bm s_0, y_0) = 0\). As an example, one
% can multiply \(b(\bm s; \bm s_0, y_0)\) with the function
% \begin{equation}
%   \label{eq:rho_b}
%   \sqrt{1 - \exp\left(-2 \|\bm s - \bm s_0\| / \rho_b\right)},
% \end{equation}
% where \(\rho_b > 0\) can be either fixed or estimated. This expression is inspired by the variance
% of a conditional bivariate Gaussian distribution with exponential correlation function.  In general,
% constraining \(\widehat Z_b(\cdot)\) by requiring that \(b(\bm s_0; \bm s_0, y_0) = 0\) is almost as
% computationally efficient as constraining by subtraction, but without any of the undesirable
% properties discussed above. In our opinion one should therefore always prefer to constrain
% \(\widehat Z_b(\cdot)\) in this way when modelling spatial conditional extremes with \texttt{R-INLA}.

\subsection{Defining \texorpdfstring{\(a(\bm s; \bm s_0, y_0)\)}{a} in \texttt{R-INLA}}
\label{sec:implementing_a}

%Based on the \texttt{R-INLA} implementation of \(b(\bm s; \bm s_0, y_0) Z(\bm s; \bm s_0)\) 
%in Section~\ref{sec:implementing_bZ}, it might seem reasonable to represent
%the entire conditional extremes model using the SPDE approximation
%\[
%  \widehat Z_{a, b}(\bm s; \bm s_0, y_0) = \sum_{i = 1}^M \phi_i(\bm s) \left( a(\bm s_i; \bm s_0, y_0)
%    + W_i b(\bm s_i; \bm s_0, y_0) \right)
%\]
%However, this approximation has a mean equal to
%\[
%  \text{E}\left[\widehat Z_{a, b}(\bm s; \bm s_0, y_0)\right] = \sum_{i, j = 1}^M \phi_i(\bm s) \phi_j(\bm
%  s) a(\bm s_i; \bm s_0, y_0),
%\]
%which in general is different from \(a(\bm s; \bm s_0, y_0)\). It is therefore better to represent
%the standardising function \(a(\bm s; \bm s_0, y_0)\) in \texttt{R-INLA} using a separate latent
%model component that does not depend on the triangulated SPDE mesh, thus allowing us to compute
%\(a(\bm s; \bm s_0, y_0)\) exactly at all locations \(\bm s\) of interest.

All components of the latent Gaussian field in \texttt{R-INLA} must be Gaussian random variables,
but \(a(\bm s; \bm s_0, y_0)\) is a deterministic function and not a random variable. However,
clearly, the deterministic vector \(\bm a\) can be approximated well by the Gaussian random vector
\(\bm a + \bm \epsilon\), where \(\bm \epsilon\) has zero mean and
covariance matrix \(\delta^2 \bm I\), with \(\bm I\) being the identity matrix and \(\delta^2\)
being a small, fixed marginal variance. Thus, using the \texttt{rgeneric}/\texttt{cgeneric}
framework, we can approximate any deterministic function \(a(\bm s; \bm s_0, y_0)\) with a latent
Gaussian random field with mean \(a(\bm s; \bm s_0, y_0)\) and diagonal covariance matrix
\(\delta^2 \bm I\). Here, we choose \(\delta^2 = \exp(-15)\).

\section{Robust inference using post hoc adjustments}
\label{sec:robustifying-inference}

\subsection{Adjusting posterior samples}

It is well known that all models are wrong, in the sense that the data, to a certain extent, always
deviate from the model assumptions. This is particularly true when modelling extremes, where most
models are based on imposing asymptotically justified assumptions onto finite amounts of data. It is
also particularly true when modelling high-dimensional data, because high-dimensional models often
are based on strict assumptions of (un)conditional independence and Gaussianity, in order to make
inference computationally tractable, and because the amount of misspecification naturally tends to
increase with the data size and dimensionality while keeping everything else constant. Accounting
for this misspecification should therefore be an important step in any successful modelling strategy
for high-dimensional spatial extremes.

Given \(n\) independent realisations \(\mathcal Y = \{\bm y_1, \ldots, \bm y_n\}\) of a random
vector \(\bm Y\) with true distribution \(G\), and a chosen likelihood
\(L(\bm \theta; \mathcal Y) = \prod_{i = 1}^n L(\bm \theta; \bm y_i)\), it is well known that the
maximum likelihood estimator \(\widehat{\bm \theta}\) for \(\bm \theta\) is asymptotically Gaussian,
i.e.,
\[
  \bm{\mathcal I}(\bm \theta^*)^{1/2} \left(\widehat{\bm \theta} - \bm \theta^*\right) \rightsquigarrow
  \mathcal N(\bm 0, \bm I), \text{ as } n \rightarrow \infty,
\]
under some weak regularity conditions \citep{white82_maxim_likel_estim_missp_model}, where \(\bm I\)
is the identity matrix and \(\bm \theta^*\) minimises the Kullback-Leibler divergence
\citep[KLD;][]{kullback51_infor_suffic} between \(L(\bm \theta; \cdot)\) and the likelihood of
the true distribution \(G\). Furthermore, \(\bm{\mathcal I}(\bm \theta)\) is the so-called Godambe
sandwich information matrix \citep{godambe60_optim_proper_regul_maxim_likel_estim}, i.e.,
\begin{equation}
  \label{eq:H-and-J}
  \bm{\mathcal I}(\bm \theta) = \bm H(\bm \theta) \bm J(\bm \theta)^{-1} \bm H(\bm \theta),
\end{equation}
with
\(\bm H(\bm \theta) = -\text{E}\left[\nabla_{\bm \theta}^2 \ell(\bm \theta; \mathcal Y)\right]\) and
\(\bm J(\bm \theta) = \text{Cov}\left(\nabla_{\bm \theta} \ell(\bm \theta; \mathcal Y)\right)\),
where \(\ell(\cdot) = \log L(\cdot)\) is the log-likelihood, and all expectations are taken with
respect to \(G\). If \(L(\bm \theta^*; \cdot)\) is equal to the likelihood of the true distribution
\(G\), then \(\bm J(\bm \theta^*) = \bm H(\bm \theta^*)\), and \(\bm{\mathcal I}(\bm
\theta^*)\) reduces to \(\bm H(\bm \theta^*)\).

From a Bayesian perspective, given a prior \(\pi(\bm \theta)\) and appropriate
regularity conditions, it is also known that the posterior density,
\(\pi(\bm \theta \mid \mathcal Y) \propto L(\bm \theta; \mathcal Y) \pi(\bm \theta)\),
%\[
%  \pi(\bm \theta \mid \mathcal Y) = \frac{L(\bm \theta; \mathcal Y) \pi(\bm \theta)}{\int L(\bm \theta;
%    \mathcal Y) \pi(\bm \theta) \text{d}\bm \theta}
%\]
converges asymptotically to a Gaussian density with mean \(\bm \theta^*\) and covariance matrix
\(\bm H(\bm \theta^*)^{-1}\) \citep{berk66_limit_behav_poster_distr_when,
  kleijn12_berns_von_mises_theor_missp}. As the sample size increases and the effect of the prior
distribution diminishes, credible intervals and confidence intervals should be expected to
coincide. However, if the likelihood is misspecified so that
\(\bm{\mathcal I}(\bm \theta^*) \neq \bm H(\bm \theta^*)\), then the resulting asymptotic
\((1 - \alpha)\)-credible interval differs from all well-calibrated asymptotic
\((1 - \alpha)\)-confidence intervals, and we say that they attain poor frequency
properties. \citet{ribatet12_bayes_infer_from_compos_likel} illustrate how easily a misspecified
likelihood function can lead to misleading inference through posterior intervals with poor frequency
properties.

Several approaches have been proposed for robustifying inference under a misspecified likelihood
\citep{chandler07_infer_clust_data_indep_loglik, pauli11_bayes_compos_margin_likel,
  ribatet12_bayes_infer_from_compos_likel, syring18_calib_gener_poster_credib_region}, but all these
methods are based on modifying the likelihood function before inference, which is impossible to do
within the \texttt{R-INLA} framework. However, \citet{shaby14_open_faced_sandw_adjus_mcmc} proposes
a post hoc adjustment method that properly accounts for misspecification in the likelihood by an
affine transformation of posterior samples when performing MCMC-based inference. Since this is a
post hoc adjustment method, it is possible to extend it for usage with \texttt{R-INLA}. Given a
sample \(\bm \theta\) from a posterior distribution based on a misspecified
likelihood, the adjusted posterior sample is defined as
\begin{equation}
  \label{eq:composite-adjustment}
  \bm \theta_{\text{adj}} = \bm \theta^* + \bm C(\bm \theta - \bm \theta^*),
\end{equation}
where the matrix \(\bm C\) is chosen such that the asymptotic distribution of
\(\bm \theta_{\text{adj}}\), as \(n \rightarrow \infty\), is Gaussian with mean \(\bm \theta^*\) and
covariance matrix \(\bm{\mathcal I}(\bm \theta^*)^{-1}\). This can be achieved by setting
\(\bm C = \left(\bm M_1^{-1} \bm M_2 \right)^{\top}\), where
\(\bm M_1^{\top} \bm M_1 = \bm H({\bm \theta}^*)^{-1}\) and
\(\bm M_2^{\top} \bm M_2 = \bm{\mathcal I}({\bm \theta}^*)^{-1}\).  The matrix square roots
\(\bm M_1\) and \(\bm M_2\) can be computed using, e.g., singular value decomposition.

A problem with the adjustment method of \citet{shaby14_open_faced_sandw_adjus_mcmc} is that it
distorts the contributions of the prior distribution. Using the formula for the
probability density function of a transformed random variable, one can show that the distribution of
the adjusted samples is
\[
  \pi(\bm \theta_{\text{adj}} \mid \mathcal Y) \propto
  L(\bm \theta^* + \bm C^{-1}(\bm \theta_{\text{adj}} - \bm \theta^*); \mathcal Y)
  \pi(\bm \theta^* + \bm C^{-1}(\bm \theta_{\text{adj}} - \bm \theta^*)).
\]
However, the prior distribution reflects our prior knowledge about \(\bm \theta\), and it should
ideally not be affected when adjusting for the misspecification in \(L\). If the prior is not
overly informative and the sample size is large enough, this may not matter, as the contribution of
the prior will be minimal. However, if that is not the case, we propose to additionally adjust the
prior distribution before inference as
\begin{equation}
  \label{eq:prior-adjusted}
  \pi_{\text{adj}}(\bm \theta) = \pi(\bm \theta^* + \bm C(\bm \theta - \bm \theta^*)) \cdot |\bm C|,
\end{equation}
such that the adjusted posterior samples have distribution
\[
  \pi(\bm \theta_{\text{adj}} \mid \mathcal Y) \propto
  L(\bm \theta^* + \bm C^{-1}(\bm \theta_{\text{adj}} - \bm \theta^*); \mathcal Y)
  \pi(\bm \theta_{\text{adj}}).
\]
Using the \texttt{rgeneric}/\texttt{cgeneric} framework, one can easily define a model in
\texttt{R-INLA} with the adjusted prior distribution \(\pi_{\text{adj}}(\bm \theta)\).

As mentioned in Section~\ref{sec:composite_likelihood}, composite likelihoods are not valid
likelihood functions. However, the theory in this section holds true for a wide class of loss
functions that includes negative composite log-likelihoods. Thus, we can still perform the
adjustment method if we exchange \(\ell(\bm \theta; \mathcal Y)\) with the composite log-likelihood
in~\eqref{eq:conditional-composite-likelihood}, and \(\pi(\bm \theta \mid \mathcal Y)\) with the
pseudo-posterior distribution
\(\pi_c(\bm \theta \mid \mathcal Y) \propto L_c(\bm \theta; \mathcal Y) \pi(\bm \theta)\), where
\(L_c(\cdot) = \exp(\ell_c(\cdot))\).

\subsection{Estimating \texorpdfstring{\(\bm C\)}{C} and \texorpdfstring{\(\bm \theta^*\)}{the KLD minimiser}}
\label{sec:adjustment_in_practice}

Here, we detail how to estimate the KLD minimiser \(\bm \theta^*\) and the matrix \(\bm C\) when
inference is based on the composite log-likelihood
in~\eqref{eq:conditional-composite-likelihood}. Our approach can also be applied in settings where
inference is based on valid likelihood functions or other types of loss functions.

The KLD minimiser \(\bm \theta^*\) can be estimated by the mode of the posterior
\(\pi_c(\bm \theta \mid \mathcal Y)\), denoted \(\hat{\bm \theta}^*\). The estimator
\(\hat{\bm \theta}^*\) indeed provides a good approximation to \(\bm \theta^*\) if the prior is not
overly informative and the sample size is large enough, but other estimators, such as the maximum
likelihood estimator, may be more suitable if this does not hold.

In order to estimate \(\bm C\) one must first estimate \(\bm H(\bm \theta^*)\) and
\(\bm J(\bm \theta^*)\). The first step towards estimating \(\bm J(\bm \theta^*)\) is to compute the
gradients \(\nabla_{\bm \theta} \ell(\hat{\bm \theta}^*; \bm y_{i, -j} \mid y_i(\bm s_j))\) for all
\(j = 1, 2, \ldots, m\) and \(i = 1, 2, \ldots, n\), such that \(y_i(\bm s_j) > t\). These gradients
can be computed analytically or estimated using numerical derivation methods. We then estimate
\(\bm J(\bm \theta^*)\) with
\begin{equation}
  \label{eq:J}
  \hat{\bm J}(\hat{\bm \theta}^*) = \sum_{i, j} I(y_i(\bm s_j) > t) \nabla_{\bm \theta}
  \ell(\hat{\bm \theta}^*; \bm y_{i, -j} \mid y_i(\bm s_j))
  \sum_{(i', j') \in \Delta(i, j)} I(y_{i'}(\bm s_{j'}) > t)  \left(\nabla_{\bm \theta}\ell(\hat{\bm \theta}^*; \bm
    y_{i', -j'} \mid y_{i'}(\bm s_{j'}))\right)^{\top},
\end{equation}
where \(\Delta(i, j)\) is the set of neighbours of \((i, j)\), i.e.,
\((i', j') \in \Delta(i, j)\) if and only if
\(\nabla_{\bm \theta} \ell(\cdot; \bm y_{i', -j'} \mid y_{i'}(\bm s_{j'}))\) is correlated with
\(\nabla_{\bm \theta} \ell(\cdot; \bm y_{i, -j} \mid y_{i}(\bm s_j))\). Summing over all non-correlated
pairs of tuples introduces unnecessary noise that could cause the estimator to approximately
equal zero \citep{lumley99_weigh_empir_adapt_varian_estim}. In practice one can often
compute~\eqref{eq:J} using a sliding window approach. This is an improvement
over the proposed estimation methods of \citet{shaby14_open_faced_sandw_adjus_mcmc}, which require
either that all log-likelihood terms are independent or that it is possible to simulate data from
the true and unknown distribution of the data.

Estimation of \(\bm H(\bm \theta^*)\) is often easier than that of \(\bm J(\bm \theta^*)\),
since the former is a matrix of expected values, whereas the latter is a covariance matrix. The law of large
number implies that, for \(n\) and \(m\) large enough, a good estimator for \(\bm H(\bm \theta^*)\)
is simply
\(\hat{\bm H}(\hat{\bm \theta}^*) = - \nabla^2_{\bm \theta} \ell_c(\hat{\bm \theta}^*; \mathcal Y)\).
%\begin{equation}
%  \label{eq:H}
%  \hat{\bm H}(\hat{\bm \theta}^*) = - \nabla^2_{\bm \theta} \ell_c(\hat{\bm \theta}^*; \mathcal Y).
%\end{equation}
Thus, all we need for estimating \(\bm H(\bm \theta^*)\) is to compute the Hessian of the composite log-likelihood
terms at \(\hat{\bm \theta}^*\).
For users of \texttt{R-INLA}, this is especially simple, since the program returns the inverse of
the Hessian matrix
\[
  \tilde{\bm H}(\hat{\bm \theta}^*) =  - \nabla^2_{\bm \theta} \pi_c(\hat{\bm \theta}^* \mid \mathcal Y) =
  - \nabla^2_{\bm \theta} \ell_c(\hat{\bm \theta}^*; \mathcal Y)
  - \nabla^2_{\bm \theta} \log \pi(\hat{\bm \theta}^*).
\]
Thus, if the prior is not overly informative and the sample size is large enough, we can set
estimate \(\bm H(\bm \theta^*)\) with \(\tilde{\bm H}(\hat{\bm \theta}^*)\). If this is not the case, we can
still estimate \(\bm H(\bm \theta^*)\) by simply subtracting the contribution of the prior distribution
from \(\tilde{\bm H}(\hat{\bm \theta}^*)\).

If we adjust the prior distribution as in~\eqref{eq:prior-adjusted}, it may be
necessary to run \texttt{R-INLA} twice: once for estimating \(\bm \theta^*\) and \(\bm C\), and once
for performing inference with the adjusted prior.

A small numerical example is shown in the supplementary material, for demonstrating that the
adjustment method is able to recover the frequency properties of a posterior distribution that is
based on a misspecified likelihood.

\section{Simulation study}
\label{sec:simulation}

We now conduct a simulation study to demonstrate our proposed workflow for modelling spatial
extremes.  Given a set of extreme realisations from simulated data we show how to compute relevant
statistics of the data and how to use these for making an informed decision about the appropriate
models for the standardising functions \(a(\bm s; \bm s_0, y_0)\) and \(b(\bm s; \bm s_0,
y_0)\). Then, we discuss details on how to define the SPDE mesh and on performing inference with
\texttt{R-INLA} and the composite likelihood. Finally, we adjust the posterior distribution for
possible misspecification and we evaluate the performance of the model fit.

We sample \(n = 10^4\) realisations of a spatial Gaussian random field
\(\mathcal Y = \{Y_i(\bm s): i = 1, \ldots, n, \bm s \in \mathcal S\}\), observed on a regular grid
\(\mathcal S\) with resolution \(1 \times 1\) and size \(100 \times 100\). The spatial Gaussian
random field has a Matérn covariance function \eqref{eq:matern} with parameters \(\sigma^2 = 1\),
\(\nu = 1\) and \(\rho = 40\), and an additional nugget effect with variance \(0.1^2\). All the
samples are created using an SPDE approximation. In order to model threshold exceedances with the
spatial conditional extremes model, we transform the observations to have Laplace marginals using
the probability integral transform. We then choose a threshold \(t\) equal to the \(99.9\%\)
quantile of the Laplace distribution.

\begin{figure}[t!]
  \centering
  \includegraphics[width=.8\linewidth]{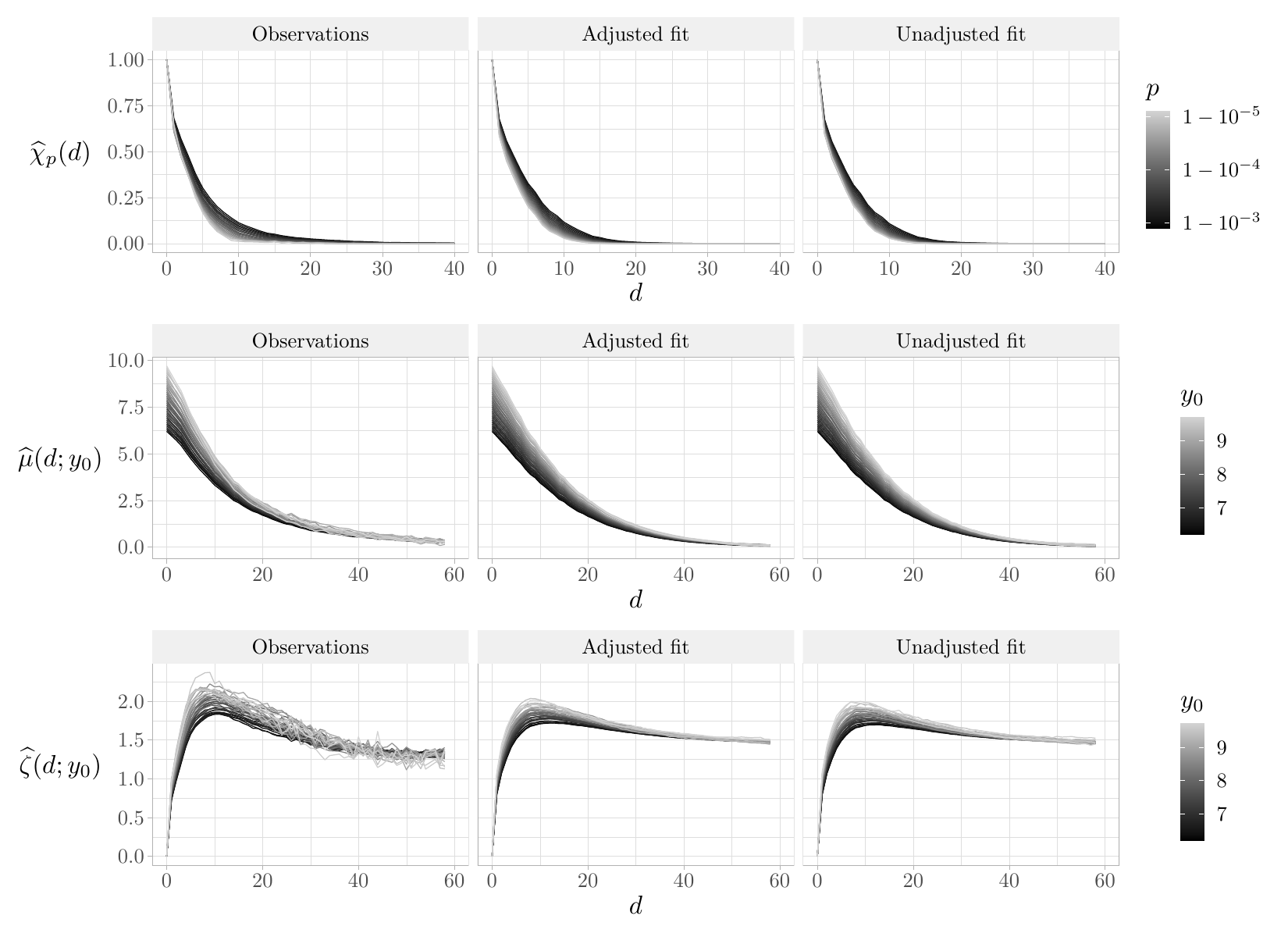}
  \caption{Empirical estimators of \(\chi_p(d)\), \(\mu(d; y_0)\) and \(\zeta(d; y_0)\) (top to bottom) from three
    different data sources. The leftmost column displays empirical estimators using the original data,
  while the two rightmost columns displays empirical estimators using data simulated from the
  adjusted and the unadjusted model fits, respectively.}
  \label{fig:simulation_properties}
\end{figure}

As a first step, we examine extremal dependence in the available data. If we (correctly) assume
stationarity and isotropy, we can denote the extremal correlation coefficient as
\(\chi_p(\bm s_1, \bm s_2) \equiv \chi_p(d)\), where \(d = \|\bm s_1 - \bm s_2\|\). We estimate
\(\chi_p(d)\) empirically using a sliding window approach, i.e., for any value of \(d\), we iterate
over all location pairs \((\bm s, \bm s') \in \mathcal S^2\) satisfying
\(|d - \|\bm s - \bm s'\|| < \delta\), for some small tolerance \(\delta > 0\), and then we count
the number of times that \(Y(\bm s) > F^{-1}(p)\) given that \(Y(\bm s') > F^{-1}(p)\), where
\(F^{-1}(\cdot)\) is the quantile function of the Laplace distribution. We here choose
\(\delta = 0.5\). Estimators for \(\chi_p(d)\) are displayed in the top-left subplot of
Figure~\ref{fig:simulation_properties}. Since the data have a Gaussian copula, we know that
\(\chi(d) = 0\) for all \(d > 0\), meaning that \(\chi_p(d)\) is far away from its limit \(\chi(d)\)
at small distances. Even if \(\chi(d)\) is unknown in practice, we can here observe a clear trend of
weakening dependence at increasing threshold levels, implying that the limit has not yet been
reached. This demonstrates the need for a model that allows for flexible modelling of sub-asymptotic
dependence, such as the spatial conditional extremes model.

To perform inference with the spatial conditional extremes model
from~\eqref{eq:conditional-model-simpson}, we must decide upon models for \(a(\bm s; \bm s_0, y_0)\)
and \(b(\bm s; \bm s_0, y_0)\). The limiting forms of these functions as \(t \rightarrow \infty\)
are already known for a spatial Gaussian random field
\citep{wadsworth22_higher_spatial_extrem_singl_condit}. However, we here assume that the
distribution of the data is unknown. Additionally, since we have chosen a finite threshold \(t\)
where \(\chi_p(d)\) is far away from its limit \(\chi(d)\), other models for \(a(\cdot)\) and
\(b(\cdot)\) may fit the data better than the known limiting forms. To examine the shape of the
standardising functions, we (correctly) assume stationarity in the sense that all model parameters
are independent of the choice of conditioning sites, and we assume that \(a(\bm s; \bm s_0, y_0)\)
and \(b(\bm s; \bm s_0, y_0)\) only depend on the distance \(d = \|\bm s - \bm s_0\|\) and threshold
exceedance \(y_0\), meaning that we can define the standardising functions as \(a(d; y_0)\) and
\(b(d; y_0)\) analogously. With these assumptions, we can visualise the forms of \(a(d; y_0)\) and
\(b(d; y_0)\) by empirically computing conditional means and variances of the data. In our model,
all random variables with distance \(d\) from \(\bm s_0\) have conditional mean
\(\mu(d; y_0) = a(d; y_0)\) and conditional variance
\(\zeta^2(d; y_0) = \sigma^2(d) b^2(d; y_0) + \tau^{-1}\), where \(\sigma^2(d)\) is the variance of
the residual field at distance \(d\) from the conditioning site, and \(\tau\) is the precision of
the nugget effect. Similarly to \(\widehat \chi_p(d)\), the empirical conditional moments of the
data can be computed using a sliding window approach. However, this time, the window must slide over
both values of \(d\) and \(y_0\). We choose a rectangular window with a width of \(1\) in the
\(d\)-direction and a width of \(0.1\) in the \(y_0\)-direction.  The conditional moment estimators
are displayed in the leftmost column of Figure~\ref{fig:simulation_properties}. The conditional
mean, \(\widehat \mu(d; y_0)\), is equal to \(y_0\) at \(d = 0\), and then seems to decay
exponentially towards zero as \(d\) increases. This fits well with the proposed model
in~\eqref{eq:a_wadsworth} if we set \(\Delta = 0\). The conditional variance is zero at \(d = 0\),
and then it increases as we move away from the conditioning site and towards ``the edge of the
storm''. Here, \(\zeta(d; y_0)\) is at its largest, as it is uncertain if observations are ``inside
the storm'', i.e., extreme, or ``outside the storm'', i.e., non-extreme. This is also where
\(\zeta(d; y_0)\) varies the most as a function of \(y_0\). Moving further away from the
conditioning site, \(\zeta(d; y_0)\) decreases to a constant, as we are certainly ``outside the
storm'', so the variance should not depend on \(y_0\) anymore. This fits well together with a model
where \(b(d; y_0) = y_0^{\beta(d)}\) and where \(\beta(d)\) decays to zero as the distance
increases. We choose to follow \citet{richards22_model_extrem_spatial_aggreg_precip} in assuming
that \(\beta(d) = \beta_0 \exp(-(d / \lambda_b)^{\kappa_b})\), with \(0 < \beta_0 < 1\) and
\(\lambda_b, \kappa_b > 0\).
%Finally, we assume that the variance parameter \(\sigma^2\) of
%\(Z(\cdot)\) is constant, meaning that the behaviour of \(\zeta(d; \cdot)\) is fully explained by
%the function \(\beta(d)\) and the constraining of the residual field that gives
%\(Z(\bm s_0; \bm s_0) = 0\) almost surely.

\begin{figure}
  \centering
  \includegraphics[width=.4\linewidth]{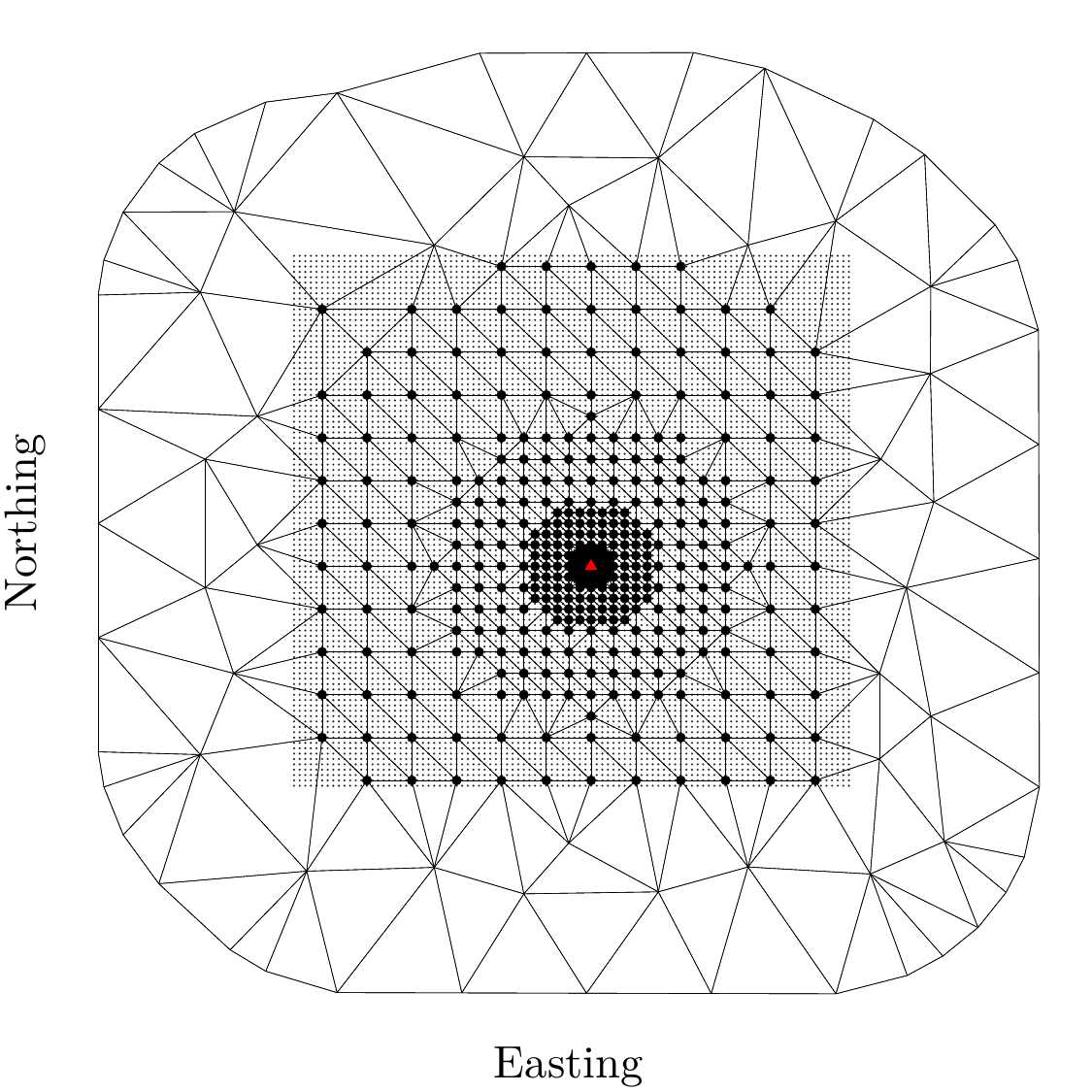}
  \caption{Given a conditioning site \(\bm s_0\) (displayed with (\textcolor{red}{\(\blacktriangle\)})),
    locations used for inference are displayed as big black dots \((\bullet)\) and
    locations in \(\mathcal S\) that are not used for inference are displayed
    as small dots \((\cdot)\).
    The SPDE mesh is displayed using black lines.}
  \label{fig:design-of-experiment}
\end{figure}

As seen in Figure~\ref{fig:simulation_properties}, the largest changes in \(\mu(d; \cdot)\) and
\(\zeta(d; \cdot)\) seem to occur when \(d\) is small. However, the majority of locations in
\(\mathcal S\) are located far away from \(\bm s_0\). To account for this and give more weight to
close-by locations, we discard some of the observations far away from \(\bm s_0\) during inference,
which also leads to increased inference speed. Figure~\ref{fig:design-of-experiment} shows an
example of the locations used to perform inference for one specific conditioning site. We stress
that these locations can vary for each conditioning site used during inference.

The SPDE approach for modelling \(Z_b(\cdot)\) requires that we define a triangulated mesh. Our
proposed constraining method from Section~\ref{sec:implementing_bZ} requires that a mesh node is
located at each conditioning site used for inference. Furthermore, the mesh should be quite dense
close to the conditioning sites to correctly capture the changes in \(b(\cdot)\). Therefore, we
define the mesh so that a mesh node is placed at each location used for inference. This can be
problematic when performing inference with a composite likelihood that depends on multiple
conditioning sites, meaning that the mesh has to be dense ``everywhere'' in \(\mathcal S\), which
leads to computationally demanding inference. Consequently, we choose to model \(Z_b(\cdot)\) with a
different mesh for each conditioning site used in the composite likelihood. Modelling different
realisations of a random field with different mesh designs is not a readily available option in
\texttt{R-INLA}, but this can be easily implemented using the \texttt{rgeneric}/\texttt{cgeneric}
framework. An example of a mesh design for one specific conditioning site is displayed in
Figure~\ref{fig:design-of-experiment}.

\begin{table}
  \centering
  \caption{Prior distributions for all model parameters.  \(\mathcal N(\mu, \sigma^2)\) denotes the
    Gaussian distribution with mean \(\mu\) and variance \(\sigma^2\). We give \(\tau\) a penalised
    complexity (PC) prior
    such that P\((\tau^{-1/2} > 1) = 0.95\). Additionally, \(\rho\) and \(\sigma\) are given the
    joint PC prior of \citet{fuglstad19_const_prior_that_penal_compl} such that
    P\((\rho < 60) = 0.95\) and P\((\sigma > 4) = 0.05\).}
  \label{tab:priors}
  \begin{tabular}{p{.35\textwidth}p{.35\textwidth}l}
  \hline
    \(\tau \sim \text{PC}(1, 0.95)\), &
    \(\log(\lambda) \sim \mathcal N(3, 4^2)\), &
    \(\log(\kappa) \sim \mathcal N(0, 3^2)\), \\
    \(\sigma \sim \text{PC}(4, 0.05)\), &
    \(\rho \sim \text{PC}(60, 0.95)\), &
    \(\log(\frac{\beta_0}{1 - \beta_0}) \sim \mathcal N(0, 2^2)\), \\
    \(\log(\lambda_b) \sim \mathcal N(3, 4^2)\), &
    \(\log(\kappa_b) \sim \mathcal N(0, 3^2)\), \\
    \hline
  \end{tabular}
\end{table}

Our chosen models for \(a(\cdot)\) and \(b(\cdot)\) are implemented using the \texttt{cgeneric}
framework, and inference is performed with \texttt{R-INLA}. The chosen priors for all the model parameters are
described in Table~\ref{tab:priors}. The priors are weakly informative, but with quite large
variances. Using all locations in \(\mathcal S\) as conditioning sites in the composite likelihood
is computationally demanding, so we define a regular sub-grid \(\mathcal S_0\) with resolution
\(6 \times 6\) and build the composite likelihood using these \(|\mathcal S_0| = 256\) conditioning locations.
The post-hoc adjustment procedure from Section~\ref{sec:robustifying-inference} is then applied to
robustify the model fit. Due to the large amount of available data we do not find it necessary to
adjust the prior distribution as proposed in~\eqref{eq:prior-adjusted}.

\begin{figure}
  \centering
  \includegraphics[width=.85\linewidth]{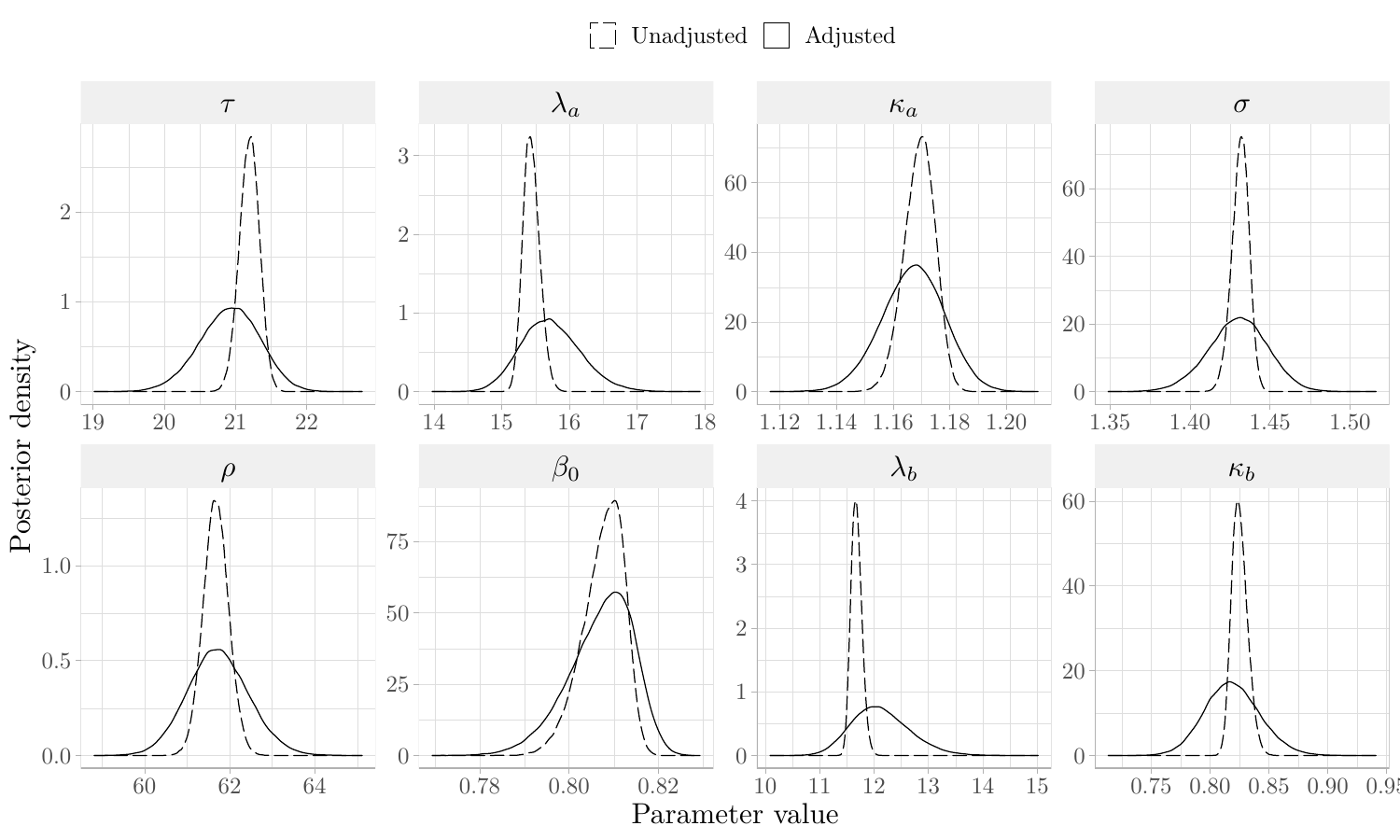}
  \caption{Posterior distributions for all model parameters from the adjusted (solid) and the
    unadjusted (dashed) model fits.}
  \label{fig:simulation_posteriors}
\end{figure}

Figure~\ref{fig:simulation_posteriors} displays the adjusted and unadjusted posterior distributions
of all model parameters. We see that the working assumption of independence in the composite
likelihood leads to overconfidence and too focused posterior distributions, and that the adjustment
method therefore increases the posterior variance to account for this misspecification.  To examine
the performance of our model fits, we simulate \(10^5\) extreme spatial fields from each fitted
model, and compute \(\widehat \chi_p(d)\), \(\widehat \mu(d; y_0)\) and \(\widehat{\zeta}(d; y_0)\)
using the simulated extremes. The estimators are displayed in the two rightmost columns of
Figure~\ref{fig:simulation_properties}. The properties of the model fits are similar to those of the
original data. There are some noticeable differences in the estimated conditional variance, which
probably stems from a too simple model for \(b(d; y_0)\). However, tailoring the perfect model
choice for \(b(d; y_0)\) is not the focus of this simulation study. Although adjusting posteriors
plays a big role in properly quantifying posterior uncertainty, there are no clear differences
between the point estimates from the two model fits in Figure~\ref{fig:simulation_properties}. This
is not very surprising, as these estimators are different types of sample means, that might be less
affected by changes in the posterior variances.

Finally, we wish to quantitatively compare the adjusted model fit with the unadjusted model fit, to
find out which one performs best.  We choose not to compare the fits by evaluating frequency
properties, as in the toy example in the supplementary material, because accurate estimation of
\(\bm \theta^*\) and the repetition of the high-dimensional simulation study hundreds of times is
too computationally demanding with our computational resources. Additionally, such comparisons are
impossible to perform for most real-life applications with finite amounts of available data.
Instead, we choose to compare the model fits by computing log-scores
\citep[e.g.,][]{gneiting07_stric_proper_scorin_rules_predic_estim} for a test data set that has not
been used during inference. Marginal composite likelihoods may be estimated using Monte Carlo
estimation: given \(n_s\) samples \(\bm \theta_1, \ldots \bm \theta_{n_s}\) from the posterior
distribution \(\pi(\bm \theta \mid \mathcal Y)\), the marginal composite likelihood for a new set of
observations \(\mathcal Y_0\) is estimated as
\(\widehat L_c(\mathcal Y_0) = \frac{1}{n_s}\sum_{i = 1}^{n_s} L_c(\bm \theta_i; \mathcal Y_0)\),
where \(L_c(\cdot)\) is the composite likelihood for the spatial conditional extremes model.  We
then denote \(\log (\widehat L_c(\mathcal Y_0))\) as the estimated log-score.  We sample
\(5 \times 10^4\) new realisations of data from the true model and locate all threshold exceedances
from the \(256\) conditioning sites used for performing inference. Log-scores are then estimated
using \(n_s = 1000\) posterior samples. This results in a log-score of \(-2502219\) for the adjusted
model fit, and \(-2504558\) for the unadjusted model fit, meaning that the adjusted model fit
attains the highest log-score, with a difference of \(2338\). Nonparametric bootstrapping of the
\(5 \times 10^4\) realisations of the spatial Gaussian random field is performed to examine if the
difference in log-score is significant. Using \(5000\) bootstrap samples, we find that the adjusted
log-score always is larger than the unadjusted log-score, with a difference between \(1000\) and
\(4500\). We conclude that the adjusted posterior performs better than the unadjusted posterior,
even though they both provide good point estimates and reasonable fits to the simulated data.

\section{Case study: Extreme precipitation in Norway}
\label{sec:case-study}

We apply our proposed methodology to the modelling of extreme hourly precipitation in Norway.
Data are presented in Section~\ref{sec:case study - data} and the inference is described in 
Section~\ref{sec:case study - inference}. Results are presented and evaluated in 
Section~\ref{sec:case study - results}.

\subsection{Data}
\label{sec:case study - data}

\begin{figure}
  \centering
  \includegraphics[width=.4\linewidth]{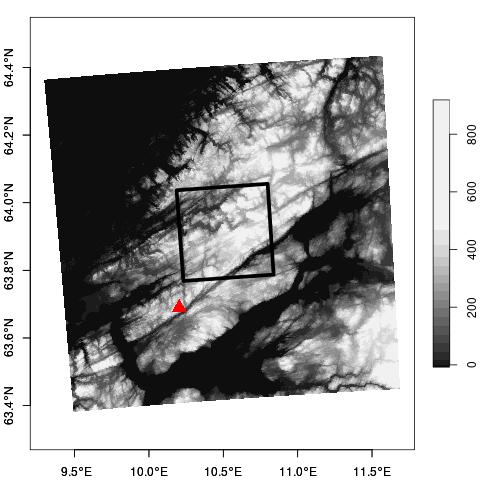}
  \caption{Elevation (m) map, over the Fosen area in central Norway. The study area \(\mathcal S\)
    is located inside the black rectangle, and the Rissa radar is displayed using a triangle
    (\textcolor{red}{\(\blacktriangle\)})).}
  \label{fig:height-map}
\end{figure}

We consider \(1 \times 1\) km\(^2\) maps of mean hourly precipitation, produced by the Norwegian
Meteorological Institute by processing raw reflectivity data from the weather radar located in Rissa
(\(63^\circ 41' 26''\)N, \(10^\circ 12' 14''\)E) in central Norway. Such maps are available online
(\url{https://thredds.met.no}), dating back to 1 January 2010.  We extract data from a rectangular
domain, close to the Rissa radar, of size \(31\times31\) km\(^2\). Denote the set of all grid points
in the rectangular domain as \(\mathcal S\). We then have \(|\mathcal S| = 961\) unique locations
containing hourly precipitation estimates. A map containing \(\mathcal S\) and the Rissa radar is
displayed in Figure~\ref{fig:height-map}. For each \(\bm s \in \mathcal S\), we extract all hourly
observations from the summer months (June, July and August) for the years 2010--2021. Removal of
missing data gives a total of 25,512 observations at each location. The number of observations with
positive precipitation amounts at each location varies from 6,000 to 17,000. These large differences
are likely numerical artefacts from the processing method of the Norwegian Meteorological
Institute. Consequently, we set all observations smaller than \(0.1\) mm precipitation equal to
0. This gives a total of between 3,500 and 4,500 positive precipitation observations at each
location.

\subsection{Modelling and inference}
\label{sec:case study - inference}

The conditional extremes model in \eqref{eq:conditional-model-simpson} is defined for a random
process with Laplace margins.  Thus, in order to perform inference with the conditional extremes
model, we standardise the marginal distributions of the precipitation data using the probability
integral transform. This is described in the supplementary materials.

\begin{figure}[t!]
  \centering
  \includegraphics[width=.8\linewidth]{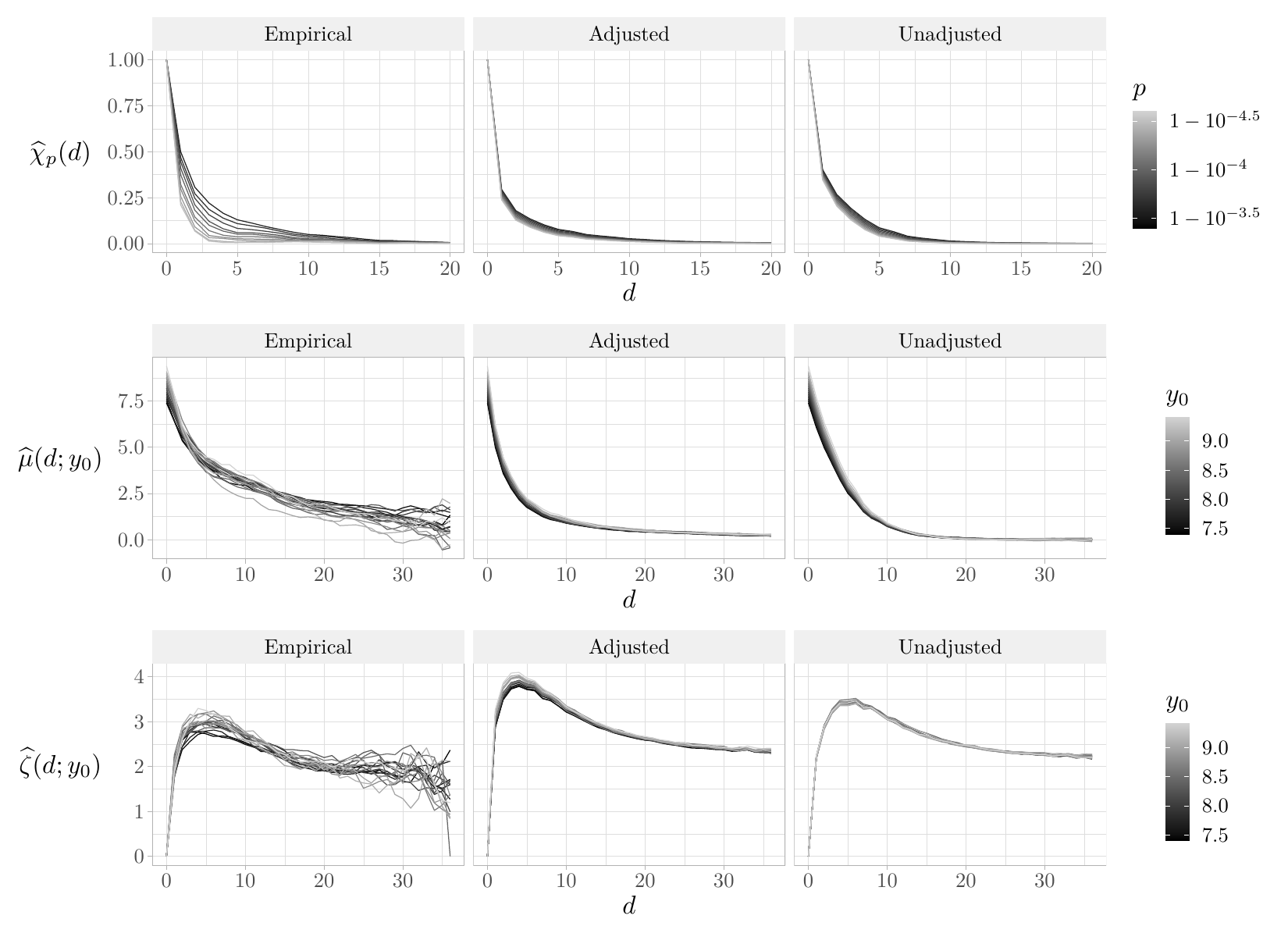}
  \caption{Empirical estimators for \(\chi_p(d)\), \(\mu(d; y_0)\) and \(\zeta(d; y_0)\) from three
    different data sources. The leftmost column displays empirical estimators using the original
    data, while the two rightmost columns displays empirical estimators using simulated data from
    the adjusted and the unadjusted model fits, respectively.}
  \label{fig:case-study_properties}
\end{figure}

Initial data exploration shows that the threshold \(t\) must be very large for a model on the form
\(a(d; y_0) = \alpha(d) y_0\) and \(b(d; y_0) = y_0^{\beta(d)}\) to provide a good
fit. Consequently, we choose a threshold equal to the \(99.97\%\) quantile of the Laplace
distribution, which yields between 0 and 5 threshold exceedances at each conditioning site. See the
supplementary materials for more details and discussion on this choice.  Estimators for
\(\chi_p(d)\), \(\mu(d; y_0)\) and \(\zeta(d; y_0)\) using this threshold are displayed in the
leftmost column of Figure~\ref{fig:case-study_properties}. Based on the lack of changes in
\(\widehat \zeta(\cdot; y_0)\) as \(y_0\) varies, we choose to model \(b(d; y_0)\) as a function not
depending on \(y_0\). We choose the model \(b(d; y_0) \equiv b(d) = 1 + b_0 \exp\left\{-\left(d /
    \lambda_b\right)^{\kappa_b}\right\}\),
with positive parameters \(b_0\), \(\lambda_b\) and \(\kappa_b\), while we use the
model~\eqref{eq:a_wadsworth} for \(a(\cdot)\), just as in Section~\ref{sec:simulation}.

With only one or two threshold exceedances at most conditioning sites, separate inference for each
conditioning site would be highly challenging. Inference is therefore performed using
\texttt{R-INLA} and the composite likelihood, based on every single conditioning site in
\(\mathcal S\). However, we remove the two last years of the data before performing inference, so
these can be used for comparing the performance of the adjusted and the unadjusted model fits.  Just
as in Section~\ref{sec:simulation}, some of the observations far away from the conditioning sites
are discarded during inference, and we also define different triangulated meshes for each
conditioning site. The Matérn smoothness parameter \(\nu\) is fixed to a value of \(1.5\). Prior
distributions are set equal to those in Table~\ref{tab:priors}, except that we exchange the
parameter \(\beta_0\) from Section~\ref{sec:simulation} with the parameter \(b_0\), where we place a
Gaussian prior on \(\log b_0\) with zero mean and a variance of \(4^2\). Finally, the post hoc
adjustment method is performed on the output of \texttt{R-INLA}. Once more, we have a large enough
sample size that we choose not to adjust the prior distribution as
in~\eqref{eq:prior-adjusted}. Estimates for \(\bm \theta^*\) and \(\bm H(\bm \theta^*)\) are
provided directly from \texttt{R-INLA}, while \(\bm J(\bm \theta^*)\) is estimated
using~\eqref{eq:J} with a sliding window that has a width of \(10\) hours.

\subsection{Results}
\label{sec:case study - results}

We simulate \(10^5\) extreme realisations from the adjusted and unadjusted model fits. Statistics of
the simulated data are displayed in the two rightmost columns of
Figure~\ref{fig:case-study_properties}. There are noticeable differences between the samples from
the two model fits. However, both model fits seem to capture a large part of the trends in the
transformed precipitation data well. Interestingly, the adjusted conditional second moments
\(\widehat \zeta(d; y_0)\) are more different from those of the original data than those of the
unadjusted model fit. However, this is not reflected in the estimated extremal correlation
coefficients, \(\widehat \chi_p(d)\). As our main goal is to capture the trends in \(\chi_p(d)\), we
see that the adjusted model fit seems to outperform the unadjusted one overall, especially for
higher values of \(p\). None of the model fits fully capture the rate of weakening dependence with
increasing thresholds, which probably requires a more complex model for \(b(d; y_0)\). As discussed
in the supplementary materials, this is outside the scope of this paper and it would require further
investigation in future research. However, any such model extension can easily be implemented using
our proposed \texttt{R-INLA} methodology.

\begin{figure}
  \centering
  \includegraphics[width=.85\linewidth]{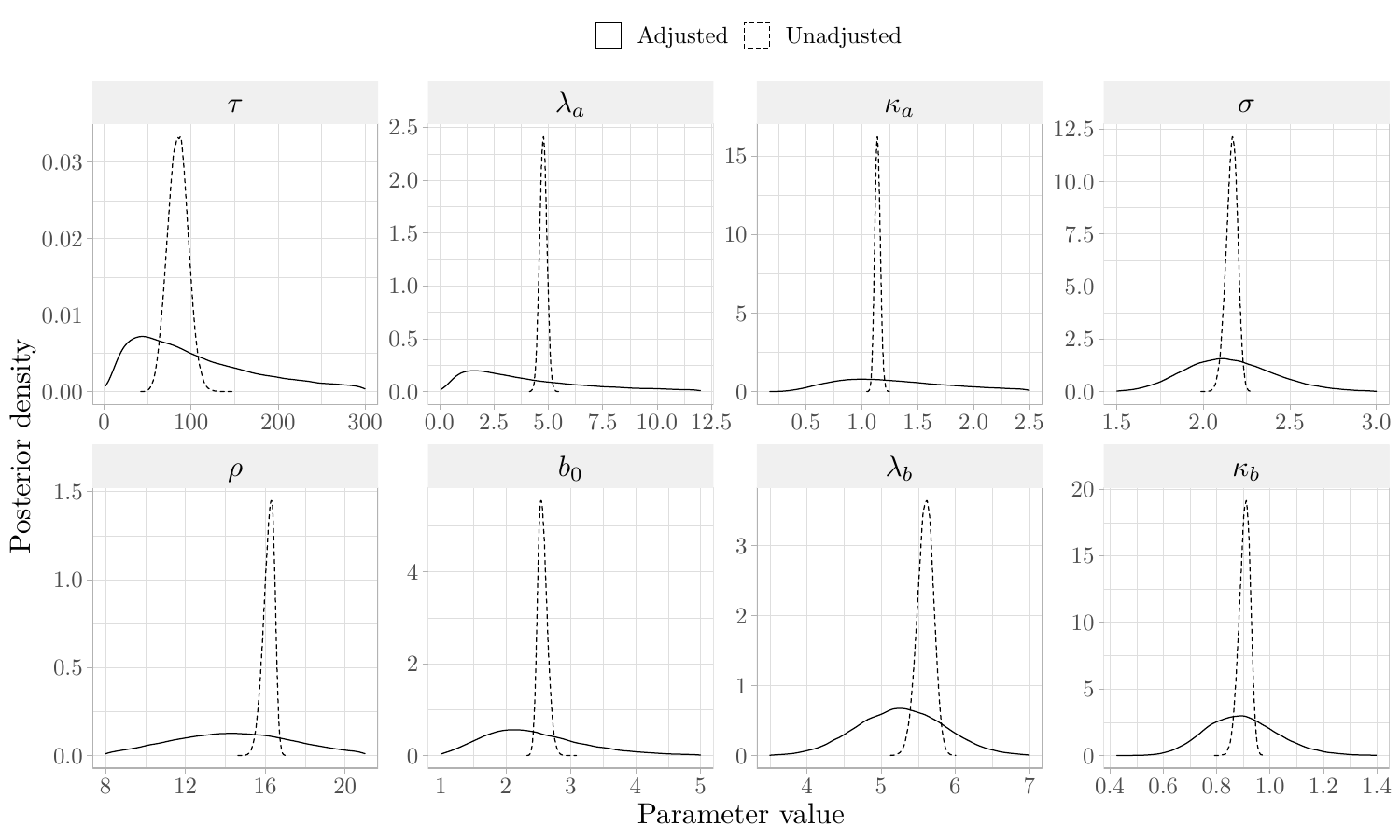}
  \caption{Posterior distributions for all model parameters of the adjusted (solid) and the
    unadjusted (dashed) model fits.}
  \label{fig:case-study_posteriors}
\end{figure}

Posterior distributions for all parameters of the two model fits are displayed in
Figure~\ref{fig:case-study_posteriors}. There are considerable differences between the adjusted and
the unadjusted posterior distributions. The latter one again too focused due to the working
assumption of independence in the composite likelihood.

To further compare the two model fits, composite log-scores are computed using the last two years of
the available data. These are estimated using \(n_s = 1000\) posterior samples. This results in a
composite log-score of \(-96520\) for the unadjusted model fit, and \(-88029\) for the adjusted
model fit, meaning that the adjusted model fit seems to perform considerably better.  Nonparametric
bootstrapping of all time points in the test data is performed to examine if the difference is
significant. Using \(5000\) bootstrap samples, we find that the difference in composite log-score is
significantly different from zero at a \(0.1\%\) significance level, so we conclude that the
adjusted model fit outperforms the unadjusted model fit.

\section{Conclusion}
\label{sec:conclusion}

We propose an efficient workflow for robust modelling of spatial high-dimensional extremes using the
spatial conditional extremes model with a composite likelihood and \texttt{R-INLA}, and a post hoc
adjustment method that corrects for possible model misspecification. The workflow is demonstrated
and shown to perform well in a large-scale simulation study, where we also propose a methodology for
selecting appropriate forms for the standardising functions \(a(\bm s; \bm s_0, y_0)\) and
\(b(\bm s; \bm s_0, y_0)\). Finally, the workflow is applied for modelling spatial high-dimensional
extremes of Norwegian precipitation data. The methodology performs well, and we are able to capture
the main extremal dependence trends in the data.

In developing our workflow, we describe a flaw in previously-used constraining methods for the
residual field in the spatial conditional extremes model, and we develop a novel constraining method
that is fast and easy to use when performing inference with \texttt{R-INLA}. We also propose and
demonstrate a general methodology for defining and implementing a large variety of spatial
conditional extremes models in \texttt{R-INLA} using the \texttt{rgeneric}/\texttt{cgeneric}
frameworks. Additionally, we propose an improved extension to the post hoc adjustment method that
allows for correct model contributions from the prior distribution.

For transforming precipitation data onto Laplace marginals, a nonparametric method is used for
estimating the marginal distributions of the original data. This method can be problematic if the
aim is to estimate properties of the original and untransformed process. Further work should
therefore focus on improving the transformation method when modelling extremes with the spatial
conditional extremes model. Additionally, even though the spatial conditional extremes model
provides good fits to the data in both the simulation study and the case study, there are still some
small differences between properties of the data and properties of the model fits. These differences
can probably be reduced by choosing better, possibly more complex, parametric or semiparametric
forms for \(a(\bm s; \bm s_0, y_0)\) and \(b(\bm s; \bm s_0, y_0)\), such as, e.g.,
\(a(\bm s; \bm s_0, y_0) = \alpha(\bm s; \bm s_0, y_0) y_0\) or
\(b(\bm s; \bm s_0, y_0) = y_0^{\beta(\bm s; \bm s_0, y_0)}\). Further work should therefore focus
on the theoretical properties of more complex models for \(a(\cdot)\) and \(b(\cdot)\), and on how
to best perform model selection with the spatial conditional extremes model. Finally, the selection
of a threshold \(t\) for the spatial conditional extremes model can have great importance for the
resulting model fit, as seen in the case study. However, to the best of our knowledge, little
attention has so far been given to the problem of threshold selection in this context. Further work
should therefore focus on methods for choosing thresholds that are large enough to provide a
somewhat correct model fit and small enough to perform inference with low uncertainty.

\section*{Acknowledgements}

The authors are grateful to Jordan Richards, Håvard Rue and Geir-Arne Fuglstad for many helpful discussions. \\

\noindent \textbf{Funding} Rapha{\"e}l Huser was partially supported by the King Abdullah University of Science and Technology (KAUST) Office of Sponsored Research (OSR) under Award No. OSR- CRG2020-4394.

\noindent \textbf{Conflict of interest} The authors report there are no competing interests to declare.

\noindent \textbf{Code and data availability} The necessary code and data for achieving these
results are available online at \url{https://github.com/siliusmv/spatialConditionalExtremes}.

\renewcommand*{\bibfont}{\footnotesize}
\printbibliography

\setcounter{section}{0}
\renewcommand{\thesection}{S\arabic{section}}% Prefix section number with S
\numberwithin{equation}{section} % Prefix equations with section number
\numberwithin{figure}{section} % Prefix figures with section number
\numberwithin{table}{section} % Prefix tables with section number

\section*{\huge Supplementary material}

\section{Post hoc adjustment toy example}
\label{sec:toy-problem}

\citet{shaby14_open_faced_sandw_adjus_mcmc} demonstrates that his proposed adjustment method is able
to properly recover the correct frequency properties of the posterior distribution. Here, we show
that the same holds when extending the post hoc method for adjusting model fits from \texttt{R-INLA},
by examining posterior frequency properties after modelling a spatial Gaussian random field with an
SPDE approximation of low rank.

Inside the spatial domain \(\mathcal S = [0, 25] \times [0, 25]\) we sample \(n\) independent
realisations of a spatial Gaussian random field with a Matérn covariance function,
which we observe at 400 random locations. The Matérn covariance function is
\begin{equation}
  \label{eq:matern-suppl}
  \text{Cov}(Z(\bm s), Z(\bm s')) = \frac{\sigma^2}{2^{\nu - 1}\Gamma(\nu)}(\kappa \|\bm s - \bm
  s'\|)^\nu K_\nu(\kappa \|\bm s - \bm s'\|),
\end{equation}
where \(\sigma^2\) is the marginal variance, \(\nu > 0\) is the smoothness parameter and
\(\rho = \sqrt{8 \nu} / \kappa\) is the range parameter of \(Z(\bm s)\). Furthermore, \(K_\nu\) is
the modified Bessel function of the second kind and order \(\nu\).  Our spatial Gaussian random
field has variance parameter \(\sigma^2 = 1\), range parameter \(\rho = 12\) and known smoothness
parameter \(\nu = 1.5\). We also add a Gaussian nugget effect with a precision of \(\tau = 100\) to
the random field.  Parameter estimation is then performed using an SPDE approximation of low rank,
i.e., based on a coarse triangulated mesh used to discretise the spatial domain. Such low-rank
approximations are typically unable to capture all the variability in the data, which means that the
nugget effect has to explain a large percentage of the variance, leading to underestimation of the
precision \(\tau\). Thus, we expect the KLD minimiser \(\bm \theta^*\) to be different from the true
parameters \(\bm \theta = (\tau, \rho, \sigma)^T\). To estimate the unknown KLD minimiser
\(\bm \theta^*\), we simulate \(n = 10^4\) realisations of the Gaussian Matérn field and compute the
maximum likelihood estimator for the misspecified SPDE model. This gives
\(\bm \theta^* = (\tau^*, \rho^*, \sigma^*) \approx (13.0, 14.5, 1.2)^T\). As expected, \(\tau\) is
severely underestimated, while \(\rho\) and \(\sigma\) are slightly overestimated.

For examination of frequency properties, we then sample \(n = 200\) new realisations of the spatial
field, and perform Bayesian inference using \texttt{R-INLA}.  We assign \(\tau\) a gamma prior with
shape 1 and scale \(2 \times 10^4\), while \(\rho\) and \(\sigma\) are given a joint penalised
complexity (PC) prior \citep{simpson17_penal_model_compon_compl,
  fuglstad19_const_prior_that_penal_compl}, setting P\((\rho<12)=0.5\) and
P\((\sigma>1)=0.5\). Inference is performed, the posterior distribution is adjusted as described in
Section~4.2 of the main paper, and credible intervals are created for both the adjusted
and the unadjusted model fits. For this simple toy example, we do not focus on adjusting the prior
distribution as described in Section~4.1. We repeat this procedure 300 times, each
time sampling \(n = 200\) new realisations which we observe at the same \(400\) locations. Coverage
frequencies can then be evaluated by examining how many of the 300 credible intervals include the
KLD minimiser \(\bm \theta^*\).

\begin{table}
  \centering
  \caption{Coverage percentages for unadjusted and adjusted credible intervals using the SPDE
    approach with a coarse mesh.}
  \label{tab:SPDE-simulation}
  \begin{tabular}{l|cccccc}
    Aim & $\tau$ & $\tau_\text{adj}$ & $\rho$ & $\rho_\text{adj}$ & $\sigma$ & $\sigma_\text{adj}$ \\
    \hline
    $90\%$ & $48\%$ & $93\%$ & $91\%$ & $90\%$ & $90\%$ & $90\%$ \\
    $95\%$ & $55\%$ & $97\%$ & $95\%$ & $95\%$ & $95\%$ & $96\%$ \\
    $99\%$ & $69\%$ & $99\%$ & $99\%$ & $98\%$ & $100\%$ & $99\%$ \\ 
  \end{tabular}
\end{table}

Table~\ref{tab:SPDE-simulation} displays the estimated coverage probabilities detailing how often the
parameters of \(\bm \theta^*\) are included in their respective credible intervals. The adjustment
of the posterior yields a considerable improvement for \(\tau\).  The unadjusted frequency
properties of \(\rho\) and \(\sigma\), however, are already good, and our adjustment method does not
deteriorate the credible intervals for these parameters.

\section{Case study prerequisites}

In order to perform inference with the conditional extremes model for a random process \(X(\bm s)\),
one must first standardise it to a random process \(Y(\bm s)\) with Laplace margins. This is
performed using the probability integral transform \citep{keef13_estim_condit_distr_multiv_variab}:
\[
  Y(\bm s) =
  \begin{cases}
    \log \left\{2 F_{X(\bm s)}(X(\bm s))\right\}, & X(\bm s) < F_{X(\bm s)}(1/2) \\
    -\log \left\{2 \left[1 - F_{X(\bm s)}(X(\bm s))\right]\right\}, & X(\bm s) \geq F_{X(\bm s)}(1/2),
  \end{cases}
\]
where \(F_{X(\bm s)}\) is the marginal distribution function of the random variable \(X(\bm s)\).
We estimate the marginal distribution functions as the site-wise empirical distribution function
of \(X(\bm s)\). However, independent standardisation of data at each location can lead to an
unrealistic lack of smoothness in the transformed process \(Y(\bm s)\). Therefore, we apply a
sliding window approach for computing the empirical distribution function, where the distribution
at location \(\bm s\) is estimated as the empirical distribution function of pooled data from all
locations \(\bm s'\) such that \(\|\bm s - \bm s'\| \leq r\) for some radius \(r\). Based on
exploratory analysis we find \(r = 5\) km to yield a realistic degree of smoothness in the estimated
marginal distributions of \(X(\bm s)\) (results not shown).

A problem when modelling precipitation is that the empirical distribution has a point mass at
zero. This leads to \(Y(\bm s)\) having a truncated Laplace distribution with a point mass, which
can cause problems during inference. In order for \(Y(\bm s)\) to follow a non-truncated Laplace
distribution, we choose to remove all zeros from the process \(X(\bm s)\) and only focus on positive
precipitation. This makes us unable to model the absence of precipitation, which can lead to a slight
overestimation of return levels for spatially aggregated precipitation. However, applying the fitted
model for estimating properties of the untransformed process \(X(\bm s)\) is outside the scope of
this paper. We believe that our choice of removing all zeros and estimating marginals using
empirical distribution functions of the positive precipitation values is acceptable given the aim
of our paper. In future research, we plan to properly model precipitation intermittence by
appropriately accounting for the point mass at zero.

\begin{figure}
  \centering
  \includegraphics[width=.99\linewidth]{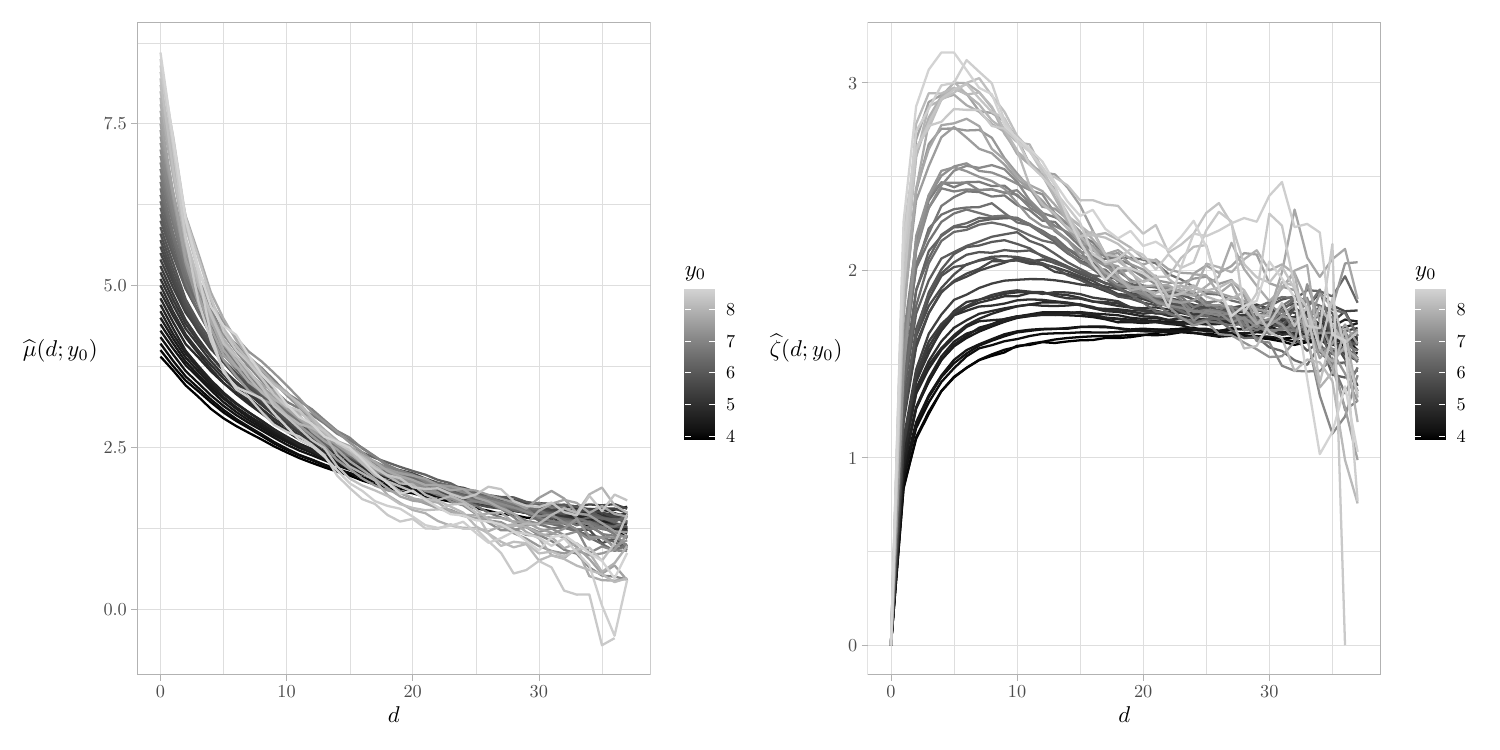}
  \caption{
    Empirical estimators for \(\mu(d; y_0)\) and \(\zeta(d; y_0)\) using the transformed
    precipitation data for \(y_0 > 4\) (\(99\%\) quantile of the Laplace distribution)}
  \label{fig:case-study_bad_properties}
\end{figure}

Similarly to the simulation study in Section~5 of the main paper, we examine extremal correlation
coefficients and empirical conditional moments of the data in order to propose a good model for the
extremes.  The spatial domain in the case study is small enough that we can assume stationarity in
the data, meaning that we can employ the same estimation methods as in the simulation
study. Empirical conditional moments of the data are displayed in
Figure~\ref{fig:case-study_bad_properties}. These estimators imply that the threshold \(t\) must be
chosen quite large for performing successful modelling with the spatial conditional extremes
model. If the threshold is chosen too low, we experience crossing in the conditional mean, i.e., for
\(y_1 \neq y_2\), \(\widehat \mu(d; y_1)\) is both smaller and larger than \(\widehat \mu(d; y_2)\)
depending on the value of \(d\). This means that a model for \(a(\cdot)\) on the form
\(a(d; y_0) = \alpha(d) y_0\) becomes unsuitable. Furthermore, there is a clear change in the shape
of the conditional variance as \(y_0\) increases, and the spread in variance at ``the edge of the
storm'' is so large that a model on the form \(b(d; y_0) = y_0^{\beta(d)}\) would require
\(\beta(d) \approx 2\) for small distances \(d\). However, \(\beta(d) > 1\) leads to an ill-defined
model \citep{wadsworth22_higher_spatial_extrem_singl_condit}. A more flexible model of the form
\(a(d; y_0) = \alpha(d, y_0) y_0\), that allows crossing, and \(b(d; y_0) = y_0^{\beta(d, y_0)}\),
that allows \(\beta(d, y_0) > 1\) for small values of \(y_0\), would probably fit well to the data,
and could easily be implemented within the \texttt{rgeneric}/\texttt{cgeneric} framework. However,
developing complex new variants of the spatial conditional extremes model is outside the scope of
this paper. Consequently, we instead choose a large threshold \(t\) equal to the \(99.97\%\)
threshold of the Laplace distribution, which removes the problems of crossing and excessively large
values of \(\beta(d)\). As we have approximately 4000 positive observations at each location, this
corresponds to a mean of \(1.2\) threshold exceedances at each conditioning site. In practice, it
yields between \(0\) and \(5\) threshold exceedances at each conditioning site.

\end{document}